\documentclass[twocolumn]{aastex631}
\usepackage{xcolor}

\newcommand{\mgii}{\ion{Mg}{2}}
\newcommand{\ovi}{\ion{O}{6}}
\newcommand{\threshold}{$\mathrm{N = 10^{12.5}~cm^{-2}}$}

\begin{document}

\title{Figuring Out Gas \& Galaxies in Enzo (FOGGIE). XV. Examining the Spatial and Kinematic Relationship between Circumgalactic \mgii\ and \ovi\ }

\shorttitle{FOGGIE XV: Circumgalactic \mgii\ and \ovi}
\shortauthors{Ticoras et al.}

\author[0009-0004-7242-0204]{Mackenzie Ticoras}
\correspondingauthor{Mackenzie Ticoras}
\affiliation{Department of Physics \& Astronomy, 567 Wilson Road, Michigan State University, East Lansing, MI 48824} 
\affiliation{Department of Computational Mathematics, Science, \& Engineering, Michigan State University, 428 S. Shaw Lane, East Lansing, MI 48824}
\email{scottm59@msu.edu}

\author[0000-0002-2786-0348]{Brian W.\ O'Shea} 
\affiliation{Department of Computational Mathematics, Science, \& Engineering, Michigan State University, 428 S. Shaw Lane, East Lansing, MI 48824}
\affiliation{Department of Physics \& Astronomy, 567 Wilson Road, Michigan State University, East Lansing, MI 48824}
\affiliation{Facility for Rare Isotope Beams, Michigan State University, East Lansing, MI 48824, USA}
\affiliation{Institute for Cyber-Enabled Research, 567 Wilson Road, Michigan State University, East Lansing, MI 48824}

\author[0000-0001-5158-1966]{Claire Kopenhafer}
\affiliation{Institute for Cyber-Enabled Research, 567 Wilson Road, Michigan State University, East Lansing, MI 48824}

\author[0000-0003-1785-8022]{Cassandra Lochhaas}
\affiliation{Center for Astrophysics, Harvard \& Smithsonian, 60 Garden St., Cambridge, MA 02138}
\affiliation{NASA Hubble Fellow}

\author[0000-0003-1455-8788]{Molly S.\ Peeples}
\affiliation{Space Telescope Science Institute, 3700 San Martin Dr., Baltimore, MD 21218}
\affiliation{Center for Astrophysical Sciences, William H.\ Miller III Department of Physics \& Astronomy, Johns Hopkins University, 3400 N.\ Charles Street, Baltimore, MD 21218}

\author[0000-0002-7982-412X]{Jason Tumlinson}
\affiliation{Space Telescope Science Institute, 3700 San Martin Dr., Baltimore, MD 21218}
\affiliation{Center for Astrophysical Sciences, William H.\ Miller III Department of Physics \& Astronomy, Johns Hopkins University, 3400 N.\ Charles Street, Baltimore, MD 21218}

\author[0000-0001-7813-0268]{Cameron W. Trapp}
\affiliation{Center for Astrophysical Sciences, William H.\ Miller III Department of Physics \& Astronomy, Johns Hopkins University, 3400 N.\ Charles Street, Baltimore, MD 21218}

\author[0009-0000-7559-7962]{Vida Saeedzadeh}
\affiliation{Center for Astrophysical Sciences, William H.\ Miller III Department of Physics \& Astronomy, Johns Hopkins University, 3400 N.\ Charles Street, Baltimore, MD 21218}

\author[0000-0001-7472-3824]{Ramona Augustin}
\affiliation{Leibniz-Institut f{\"u}r Astrophysik Potsdam (AIP), An der Sternwarte 16, 14482 Potsdam, Germany}

\author[0000-0001-9158-0829]{Nicolas Lehner}
\affiliation{Department of Physics and Astronomy, University of Notre Dame, Notre Dame, IN 46556}

\author[0000-0002-6804-630X]{Britton D.\ Smith}
\affiliation{Institute for Astronomy, University of Edinburgh, Royal Observatory, EH9 3HJ, UK}

\author[0000-0002-2591-3792]{J. Christopher Howk}
\affiliation{Department of Physics and Astronomy, University of Notre Dame, Notre Dame, IN 46556}

\begin{abstract}
Understanding the thermodynamic properties of the circumgalactic medium (CGM) is key to uncovering the baryon cycle in galaxies. Here, we present spatial and kinematic relationships between \mgii\ and \ovi, as representatives for low- and high-ion-bearing gas, in the cosmological zoom-in galaxy simulation suite FOGGIE, a set of Milky-way-like galaxy simulations with high CGM resolution. We find the \ovi-bearing gas exists as a diffuse halo around the Galactic disk, while the \mgii-bearing gas is more centrally located. We investigate the covering fraction, probability of co-observation, cokinematic correspondence of these ions using two different analysis methods. We make both mock sightlines using two-dimensional projections of our simulations treating these cells as integrated lines of sight and we create one-dimensional ray objects and use the  Synthetic
Absorption Line Surveyor Application code \citep{boyd_salsa_2020} to investigate individual gas structures that contribute most to the line of sight column densities, which we call mock absorbers. We explore the relative kinematics of these mock absorbers and find \mgii\ and \ovi\ appear to have a cokinematic relationship when looking at absorber pairs with the closest relative velocity like in \citet{werk_cos-halos_2016}. However, this does not necessarily correspond with a close spatial separation, meaning many \ovi\ and \mgii\ absorber pairs only appear to be cokinematic but are physically unrelated. Taking a more holistic look at \mgii\ and \ovi\ absorber pairs reveals a much weaker correlation between these two ions.  
\end{abstract}



\section{Introduction} \label{sec:intro}

\begin{figure*}[htbp]
    \centering
    \includegraphics[width=\textwidth]{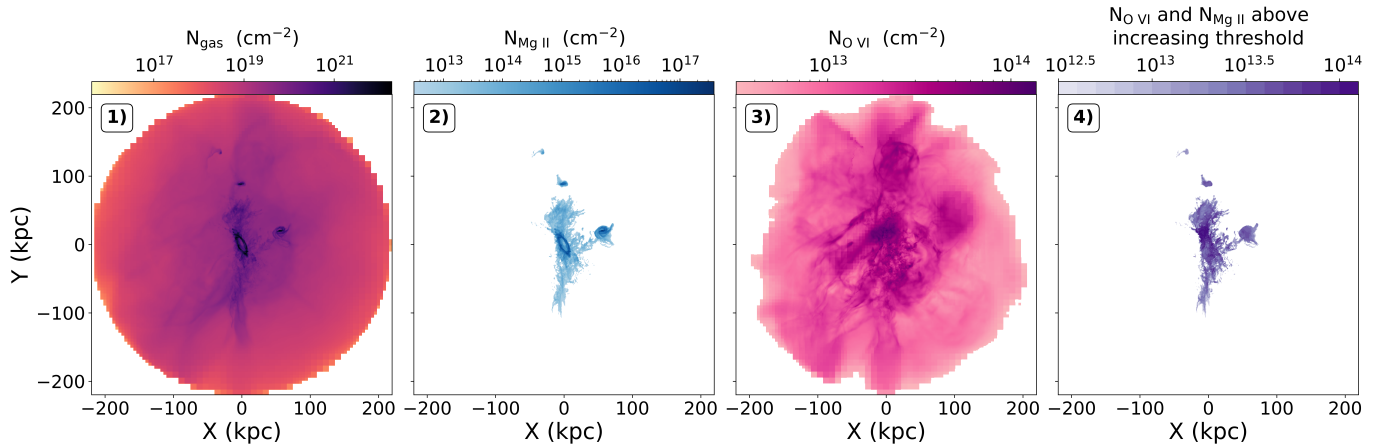}
    \caption{
    All four panels show a projection of one of the FOGGIE galaxies (Blizzard) at $\mathrm{z = 0}$. Left to right:
    1) A projection of total gas density, represented as a column density in $\mathrm{cm^{-2}}$. 
    2) A projection of \mgii\ column density with only column densities greater than the threshold of \threshold\ shown. 
    3) A projection of \ovi\ column density with only column densities greater than the threshold of \threshold\ shown. 
    4) The observable column density threshold is incremented between 
    $\mathrm{10^{12.5}~cm^{-2} < N < 10^{14.1}~cm^{-2}}$, which marks the highest column density for \ovi\ in the dataset. The pixels are colored by whether the \mgii\ and \ovi\ column densities are both above the threshold listed in the colorbar.}
    \label{fig: Covering-Fraction}
\end{figure*}

The circumgalactic medium (CGM) is a vast multiphase and ionized plasma surrounding a galaxy, which is thought to contain a significant fraction of a galaxy's total baryons \citep{werk_cos-halos_2014, lehner_evidence_2015, stocke_characterizing_2013}. Baryons move through the CGM exchanging with the galaxy disk and intergalactic medium in what is normally termed the ``baryon cycle" \citep[see, e.g.,][]{peroux_cosmic_2020}.
This cycle implicitly controls the amount of material in different galactic reservoirs (stars, interstellar medium, intergalactic medium, etc.) affecting key properties of a galaxy like star formation rate, metallicity, etc.  
Therefore, understanding the structure and dynamics of the material in the CGM is critical to understanding galaxy dynamics on many scales. 

A key way we can probe the thermodynamic structure of the CGM is through ionic column density measurements from absorption spectra.
The measurements take place by looking at the light from quasi-stellar objects (QSOs) that passes through the CGM of a galaxy. 
Light absorption occurs at certain wavelengths depending on the atomic and ion composition of the plasma. 
These absorption features are detectable in spectra and the ion composition reveals the thermodynamic properties of the medium. 
Measuring these features in spectra is a key way that we can determine the density and temperature of the CGM.  
Study of the CGM through ion column densities often occurs in population studies \citep[e.g.,][]{chen_what_2010, werk_cos-halos_2016, lehner_kodiaq-z_2022}, which can give us insight into the general properties of galaxies over a range of system masses. 
However, surveys of Andromeda and our own Milky Way CGM are also possible, allowing us to get a more detailed picture of an individual galaxy system \citep[e.g.,][]{fox_kinematics_2020, 2024ApJ...965..100Q, lehner_evidence_2015, lehner_project_2020, lehner_project_2025}.
Although limited, these types of observations can tell us about the temperature, density, and composition of the CGM as well as the kinematic structure of the gas.

We often break our collection of ion features into groups. Neutral or weakly ionized atoms (e.g., \ion{Si}{1}, \mgii) tend to inhabit colder, more dense regions of gas, whereas highly ionized atoms (\ion{C}{4}, \ovi, etc.) tend to inhabit hotter, less dense regions of gas (see Figure 6 in \cite{tumlinson_circumgalactic_2017} for more details). Here, we refer to these two groups as ``low-ions" and ``high-ions."
A broad range of both low and high-ions are often measured in a single observational sightline, and it is valuable to investigate the conditions that lead to the detection of multiple gas phases.
This work uses \mgii\ and \ovi\ to represent the low-ion and high-ion categories, respectively. 
We choose these two ions because they have strong oscillator strength \citep{tumlinson_circumgalactic_2017} and often appear in CGM observations \citep[e.g.,][]{cherrey_muse_2025, chen_what_2010, 
chen_circumgalactic_2025,
ho_kinematics_2025,
sameer_cos_2024,
dutta_musequbes_2025}. 

In general, we are also interested in the relative kinematics of ions. 
The line of sight velocity of absorption features is a quantity we can measure by looking at the wavelength offset from rest-frame absorption --- after correcting for the redshift of the host galaxy. 
Absorbers often have several components, which indicate absorption from multiple distinct structures in the CGM. 
It has been suggested in observations \citep[e.g.][]{werk_cos-halos_2016, tripp_highresolution_2008} that the line of sight velocity components of \ovi\ aligns closely with those of low-ions.
This would indicate that \ovi\ is comoving with low-ions. 
This is a non-intuitive finding since these ions occupy substantially different temperature and density phase spaces and should be uncorrelated with one another.
However, this is a complicated problem because even within a single survey the velocity offsets of \ovi\ absorbers vary widely and these features can be broad \citep{lehner_galactic_2014, werk_cos-halos_2016}.
We largely complete our analysis through the lens of \citet{werk_cos-halos_2016} using specifically \mgii\ as our tracer for low-ion-bearing gas and we directly compare to results from that paper. 

This paper traces \mgii\ and \ovi\ in the high-resolution FOGGIE simulations, which are a suite of Milky-Way like galaxy simulations with forced high resolution for much of the circumgalactic region (for more details on the simulations, see Section \ref{sec: simulations}). 
Numerical simulations like these are valuable as they allow us to probe our models and hypotheses surrounding galaxies. 
Since we are often limited in the data that can be obtained on individual galaxies (i.e., it is typical to have one QSO sightline through a galaxy), simulations allow us to synthesize observational population statistics and theoretical models to attempt to imitate a galaxy and its CGM. 
The main focus of our investigation is to examine the spatial and kinematic relationship between cool and warm gas, as traced by \mgii\ and \ovi, in the FOGGIE simulations and compare these results to observations.

Section \ref{sec: simulations} outlines the  characteristics of the simulations. 
Section \ref{sec: 2D results} discusses the spatial properties of the \mgii\ and \ovi-bearing gas with mock sightlines created by projecting the simulations into two dimensions.
Section \ref{sec:SALSA} examines the spatial structure and kinematics of individual absorbers that are computed via the SALSA package \citep{boyd_salsa_2020}, which picks out these absorbers from one-dimensional ray objects that model observational sightlines.
Section \ref{sec: discussion} investigates insights we gain by comparing the results of Section \ref{sec: 2D results} and Section \ref{sec:SALSA},  connects the results of previous sections with observations data, and discusses limitations and caveats of this work. 
Finally, Section \ref{sec: summary} summarizes the results and key points of this paper.

\section{Simulations and Methods} \label{sec: simulations}

\begin{deluxetable*}{ccccccc}
\label{tab:galaxyparams}
\tablecaption{The virial mass at $\mathrm{z=0}$ and the virial radii at relevant redshifts for the FOGGIE halos}
\tablecolumns{7}
\tablehead{
\colhead{Halo} &
\colhead{$\mathrm{M_{vir}~(10^{12}~M_\odot)}$} &
\multicolumn{5}{c}{$\mathrm{R_{vir} (kpc)}$} \\
\colhead{} & 
\colhead{z = 0} &
\colhead{z = 3} &
\colhead{z = 2} &
\colhead{z = 1} &
\colhead{z = 0.5} &
\colhead{z = 0} 
}
\startdata
 Tempest   &  $\mathrm{0.5}$ & 29.9 & 54.8 & 100.4 & 135.4 & 169.6 \\
 Maelstrom & $\mathrm{1}$    & 42.6 & 72.7 & 131.4 & 164.7 & 211.8 \\
 Squall    & $\mathrm{0.8}$  & 26.3 & 54.9 & 109.5 & 157.5 & 198.8 \\
 Blizzard  & $\mathrm{1.1}$  & 43.0 & 80.9 & 131.6 & 176.1 & 219.6\\
 Cyclone   & $\mathrm{1.6}$  & 40.2 & 92.6 & 147.6 & 200.0 & 251.5\\
 Hurricane &  $\mathrm{1.6}$ & 16.4 & 83.0 & 143.2 & 197.6 & 254.1\\
\enddata 
\end{deluxetable*}

We outline details of the ``Figuring out Gas \& Galaxies in Enzo" (FOGGIE) simulations that we use for this paper in Section \ref{subsec: FOGGIE SIMS}.
We then describe how we obtain the ionization fractions on a cell-by-cell basis in Section \ref{subsec: Ionization Extraction}.

\subsection{The FOGGIE Simulations} \label{subsec: FOGGIE SIMS}
We use the FOGGIE simulations, first introduced in \citet{peeples_figuring_2019}.
For these simulations, an initial, cosmological simulation of size $100~h^{-1}~\rm{comoving}~\rm{Mpc}$ was generated at low resolution using ENZO --- a hydrodynamical adaptive mesh refinement code \citep{bryan_enzo_2014, brummel-smith_enzo_2019}. 
These initial simulations were generated using a flat $\mathrm{\Lambda CDM}$ model of the universe with $1 - \Omega_\Lambda = \Omega_m = 0.285$, $\Omega_{\mathrm{baryon}} = 0.00461$, and $h = 0.695$.
From this, six halos were identified (Tempest, Squall, Maelstrom, Blizzard, Hurricane, and Cyclone) with \citet{simons_figuring_2020} being the first paper to discuss the properties of all six halos. 
Table \ref{tab:galaxyparams} summarizes the z=0 virial mass for each halo, as well as the virial radius at all redshifts relevant to this paper.
Additional details about these FOGGIE halos can be found in \citet{simons_figuring_2020} and \cite{wright_figuring_2024}.

We have selected these six halos such that  at $\mathrm{z=0}$ they are Milky Way-mass (between $\mathrm{M_{vir} \approx (0.5 - 1.6) \times 10^{12}~M_\odot}$) and have no major mergers (defined as a mass ratio of 10:1 or lower) after $\mathrm{z=2}$, which mirrors the believed last major merger of the Milky Way \citep{helmi_merger_2018}. 
After identifying these halos the simulations were re-run starting at $\mathrm{z=100}$ with much greater mass resolution around these halos.
The unique aspect of the FOGGIE simulations starts at $\mathrm{z=6}$ where we implement a ``forced refinement" box. 
The normal mesh refinement scheme in ENZO refines cells based on physical parameters like density, dark matter overdensity, and metallicity. However, The FOGGIE forced refinement box overrides this process requiring at least $\mathrm{\sim 1~comoving~kpc}$ $\mathrm{(n_{ref} = 9})$ resolution within $\mathrm{\pm~100~h^{-1}~comoving~kpc}$ from the galaxy center. The resolution within the box may be higher with the smallest cells being $\mathrm{274~comoving~pc}$ $\mathrm{(n_{ref} = 11})$ in size, depending on the normal refinement parameters and the local cooling length of the gas. Outside of the forced refinement box the normal ENZO refinement schemes are used.
This allows us to obtain high spatial resolution for gas extending out to a significant fraction of the virial radius. 
It has previously been shown that simulation resolution can greatly affect the simulated gas distribution in the CGM \citep{peeples_figuring_2019, corlies_figuring_2020, vandevoort_cosmological_2019, hummels_impact_2019, suresh_zooming_2019}, so modeling with high resolution is key to modeling the CGM accurately --- particularly in regions normally associated with warm/hot CGM gas, which is often less resolved than cooler, denser parts of the galaxy. 
We do still resolve pockets of cold gas in the CGM as shown in \cite{2025ApJ...993...52A} and we do not believe these features are highly dependent on the simulation resolution.

These simulations implement metallically-dependent radiative cooling and chemistry through Grackle \citep{smith_grackle_2017} with redshift-dependent UV backgrounds following \citet{haardt_radiative_2012} that include self-shielding. 
We model star formation and thermal feedback following the methods of \citet{cen_galaxy_1992, cen_where_2006} with small changes following \citet{smith_nature_2011}. We form star particles given the criteria from \citet{smith_nature_2011}, which depends on baryon overdensity, velocity divergence and cooling time.
The FOGGIE simulations create star particles with a minimum mass of $\sim\mathrm{10^3~M_\odot}$, depending on redshift. 
These star particles enrich the surrounding gas with metals produced by stellar feedback. We again use the algorithm from \citet{smith_nature_2011}, which  distributes feedback energy to the central cell and the 26 nearest cells, rather than only the central cell. 
This keeps the cells' cooling times longer and within the resolution of the simulation. As discussed in \citet{simons_figuring_2020} the cooling length is resolved for $\mathrm{>99\%}$ of the volume of these simulations, meaning that we resolve almost all of the relevant structures for this analysis.
These simulations do not include a method for active galactic nucleus feedback.

\subsection{Extracting Ionization} \label{subsec: Ionization Extraction}

In the FOGGIE simulations the metallicity is tracked as a single field, but for this analysis we are interested in decomposing the metallicity into specific ionization states. 
We use Trident \citep{TridentMethods} to add ion concentrations on a cell-by-cell basis. 
Trident calculates ion fractions by interpolating between values in the CLOUDY \citep{2013CLOUDY} tables. 
The ion fractions are based on the temperature, density, and redshift of the cell. 
These tables assume Solar abundances and assume a UV background following \citet{haardt_radiative_2012}.
More specifically, we consider the self-shielding versions of these tables from \citet{emerick_simulating_2019}, which include self-shielding by integrating to a depth into the cloud equivalent to the local Jeans length or a maximum of 100 pc. 

\begin{figure*}[htbp]
    \centering
    \includegraphics[width=\linewidth]{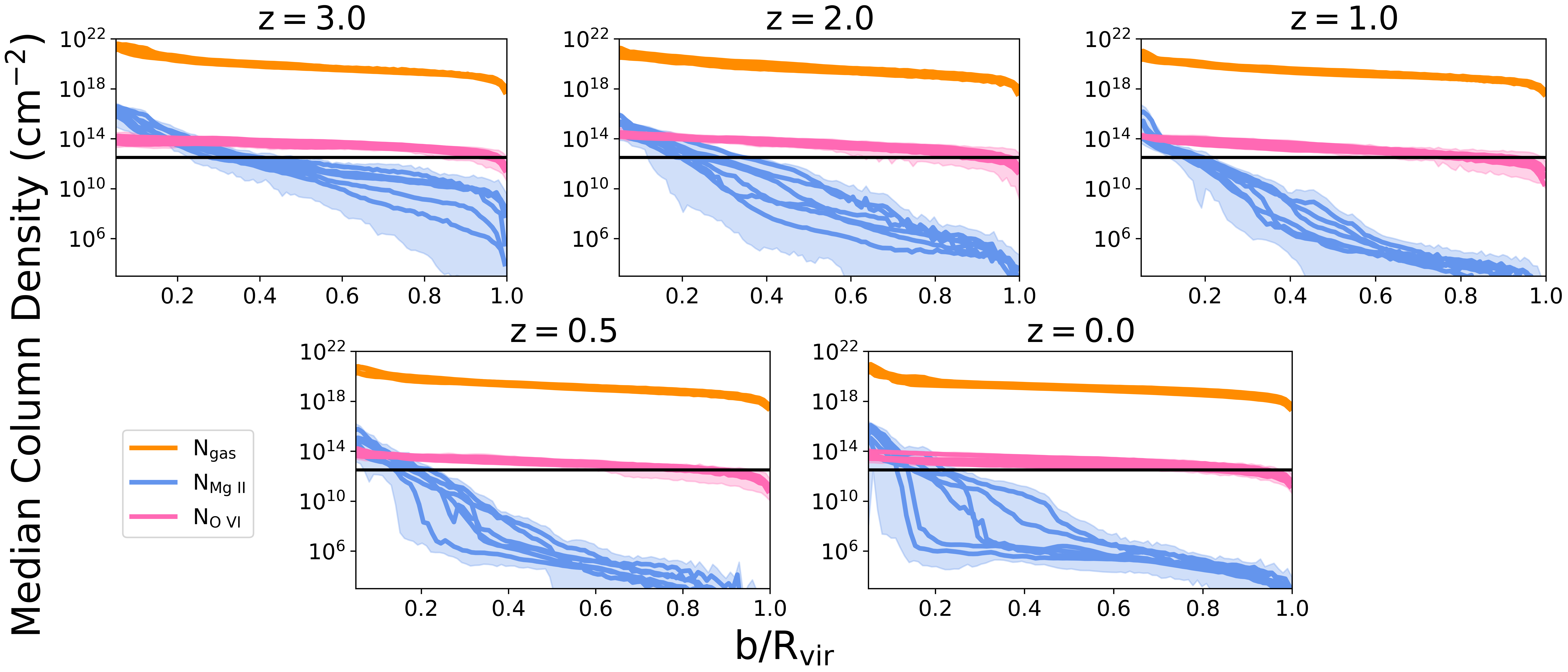}
    \caption{Median column density versus impact parameter for the two-dimensional mock sightline datasets at redshifts $\mathrm{z = 3,~2,~1,~0.5,~and~0}$. 
    We have normalized our impact parameter, b, by the virial radius, $\mathrm{R_{vir}}$.
    The orange lines represent the total gas column density, the blue line represents the \mgii\ column density, and the pink line represents the \ovi\ column density.
    Each line is the average across 10 orientations (the x, y, and z axes of the simulation and 7 random orientations) for each of the six FOGGIE halos. 
    The shaded regions represent the absolute maximum and minimum across all orientations and all halos.
    The horizontal black line is at \threshold\ which represents the observable threshold that we choose for this analysis. 
    }
    \label{fig: Covering-Fraction-Profile}
\end{figure*}

\begin{figure*}[htbp]
    \centering
    \includegraphics[width=\linewidth]{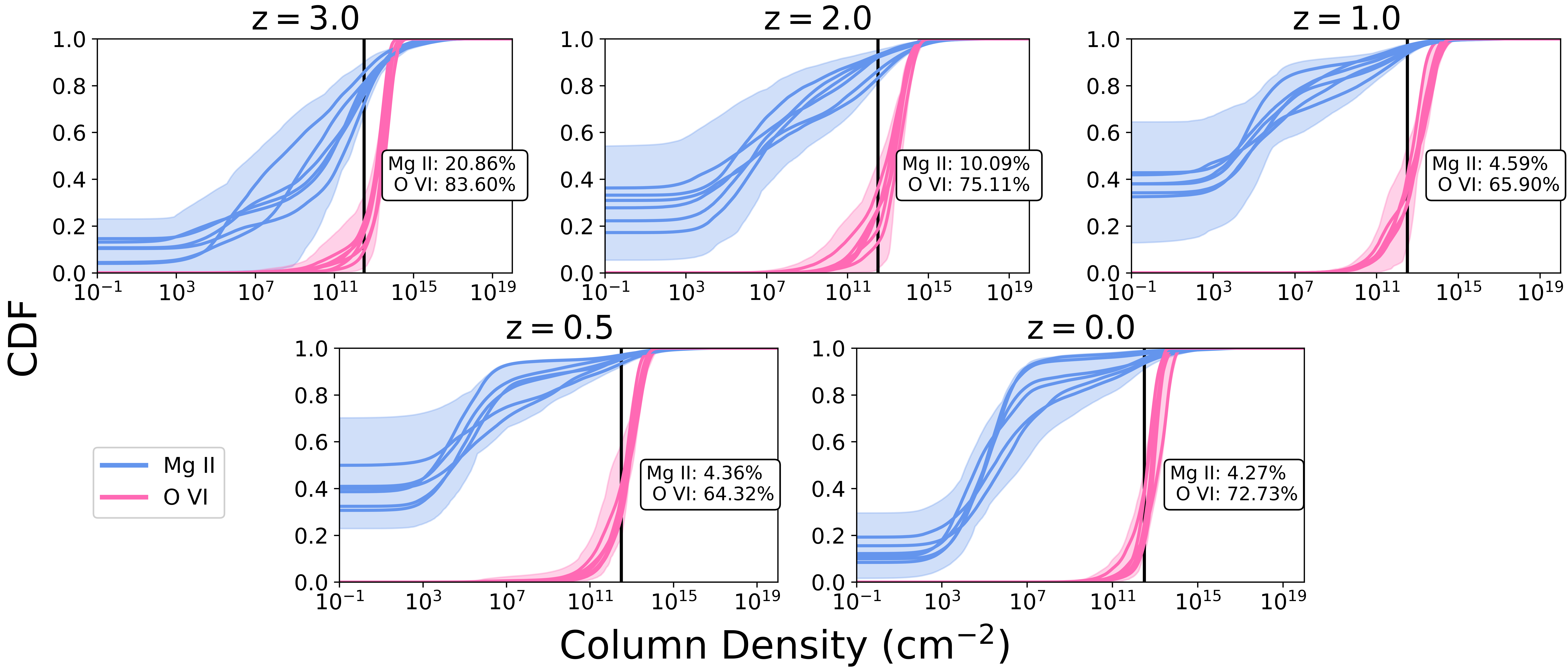}
    \caption{Cumulative distribution functions (CDF) of \mgii\ (blue) and \ovi\ (pink) column density values. 
    The black vertical line represents our observable threshold of \threshold.
    As an annotation, we report the percent of mock sightlines that contain \mgii\ and \ovi\ column densities above the threshold as averaged across all of six FOGGIE halos at each redshift.
    Each line represents the average across 10 orientations for a FOGGIE halo and the shaded regions represent the absolute maximum and minimum across all orientations for all halos. We report these distributions at redshifts of $\mathrm{z = 3,~2,~1,~0.5,~and~0}$.}
    \label{fig: CoveringFractionCalculation}
\end{figure*}

\section{Mock Sightlines In Projected Grids} \label{sec: 2D results}

\begin{figure*}[htbp]
\centering
\includegraphics[width=\linewidth,keepaspectratio]{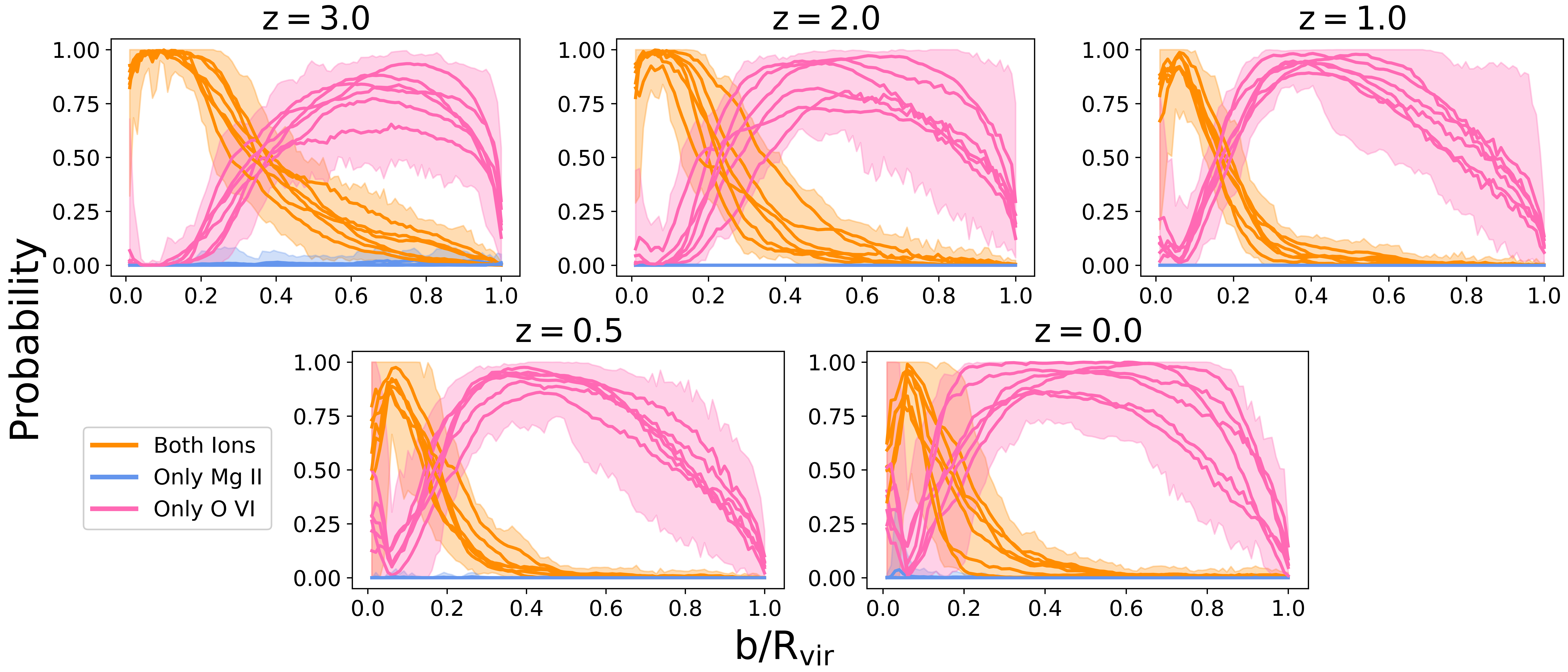}
    \caption{Probability of a projected mock sightline containing column densities of only \mgii, only \ovi, or both above the threshold of \threshold\ as a function of impact parameter.
    The orange represents the probability of both being observed, the blue represents the probability of only \mgii\ being above the threshold and the pink represents the probability of only \ovi\ being above the threshold.
    Each line represents the average across 10 orientations for a FOGGIE halo and the shaded regions represent the absolute maximum and minimum across all orientations for all halos.
    We report these distributions at redshifts of $\mathrm{z = 3,~2,~1,~0.5,~and~0}$.}
    \label{fig:ObservableFraction_vs_ImpactParameter}
\end{figure*}

For this section, we construct datasets of mock sightlines for single redshift snapshots of the FOGGIE simulations (as seen
in Figure \ref{fig: Covering-Fraction}).
We consider only cells between $\mathrm{0.05~R_{vir}}$ and $\mathrm{R_{vir}}$ for each halo at each redshift (We consider redshifts of $\mathrm{z = 3,2,1,0.5~and~0}$). 
This gives us just material inside the CGM and cuts out material from the galactic disk and intergalactic medium.
We project the three-dimensional FOGGIE simulation snapshots along a single axis only including the material between our range of $\mathrm{0.05~R_{vir}} - \mathrm{R_{vir}}$ in our projections. The result is a two-dimensional dataset upon which we impose a uniform grid of 800 × 800 cells. These cells act as our mock sightlines, which imitate QSO absorption spectroscopy as they contain information integrated along line of sight for our defined CGM region. From this grid of mock sightlines we extract column density. 
This whole process is repeated for 10 different orientations (the x, y, and z axes of the simulation and 7 random orientations). We take the average across these orientations for Figures \ref{fig: Covering-Fraction-Profile}, \ref{fig: CoveringFractionCalculation}, and  \ref{fig:ObservableFraction_vs_ImpactParameter} to limit the effect that viewing angle has on our results. 

We choose a threshold of \threshold\ for \mgii\ and \ovi\ column densities, which we consider to be approximately ``observable" column densities of these ions \citep{tumlinson_circumgalactic_2017}, so we can discuss our results in terms of the observational limits. 
We choose \mgii\ and \ovi\ to represent low-ion-bearing and high-ion-bearing gas (or cool and warm temperature gas), respectively.
These ions are often detected in spectra of the CGM of Milky Way-like galaxies. 
The complications associated with column densities observationally and the limits of setting specific thresholds are discussed in Section \ref{sec: discussion}.

Section \ref{subsec: Radial Profiles} discusses radial profiles of the \mgii\ and \ovi\ column densities.
Section \ref{subsec: covering-fraction} will discuss the probability of our sightlines containing \mgii\ and \ovi\ column being above the threshold and how this probability changes as a function of impact parameter. 

\subsection{Projected Radial Distribution of \mgii\ and \ovi}\label{subsec: Radial Profiles}

Figure \ref{fig: Covering-Fraction} shows a visual representation of these two-dimensional datasets of mock sightlines. 
The first three panels show the total, \mgii, and \ovi\ column density in a two dimensional projection of the FOGGIE galaxy ``Blizzard" at $\mathrm{z=0}$. 
The fourth panel shows a map of the overlap between \mgii\ and \ovi\ column densities as we incrementally increase the ``observability threshold" from $\mathrm{12.5 \le \log_{10}{N} \le 14.1}$. 
This shows us where there are large concentrations of both ion column densities. 
We can see in Panel 2 that \mgii\ column densities above the threshold tend to be close to the galactic center, whereas \ovi\ column densities above the threshold in Panel 3 tend to be more uniformly distributed within the virial radius. 
Additionally, Panel 4 shows that the overlap between observable column densities of \mgii\ and \ovi\ is mostly constrained by the location of observable \mgii\ sightlines. This demonstrates that observable sightlines of \ovi\ should be very common at most impact parameters. 

In Figure \ref{fig: Covering-Fraction-Profile}, we quantify this phenomenon with radial profiles of our two-dimensional datasets exploring the median of the total gas, \mgii, and \ovi\ column densities. 
Each line represents the median values over 10 different orientations (the x, y, and z axes of the simulation and 7 random orientations) for each FOGGIE halo. 
The shaded region behind the \mgii\ and \ovi\ column density profiles represents the absolute maximum and minimum found across all galaxies and orientations. 
The black horizontal line denotes our observable threshold at \threshold.
We see in this figure that the column density radial profile of \mgii\ starts higher than the \ovi\ profile, but drops below the observable threshold at around $(0.2-0.3)\mathrm{R_{vir}}$. 
The median \ovi\ column density is flat, however, remaining around $\mathrm{N \sim 10^{13}-10^{14}~cm^{-2}}$ out to nearly the virial radius. 
These trends hold across all six FOGGIE halos. 

It is important to note for Figure \ref{fig: Covering-Fraction-Profile} that the virial radius is not constant across the halos or across time. The halos have inherently different sizes and range in $\mathrm{z=0}$ mass between $\mathrm{M_{vir} \approx (0.5 - 1.5) \times 10^{12}~M_\odot}$. This results in a range of virial radii between 16--43~kpc at $\mathrm{z=3}$ which increase over time to 170--252~kpc at $\mathrm{z=0}$. 

All of the halos have a fairly tight agreement for column density values in the profiles, especially for \ovi\ column densities and the total column density. 
There appears to be more scatter in the \mgii\ column density, which could mean that \mgii\ is not equally distributed throughout the CGM but is preferentially aligned along a direction, potentially with the galactic disk.  
There is also little difference between each radial profile across redshift. 
The total and \ovi\ column densities maintain similar values and shape across redshift.  
The \mgii\ profiles appear to drop off more sharply at low redshift than they do at high redshift with the normalized impact parameter. 
However, it is important to note that the virial radius increasing is a contributing factor and the absolute size of the \mgii\ distribution stays closer to constant over time in absolute distance. 

\subsection{Probability of column densities above the threshold} \label{subsec: covering-fraction}

\begin{figure*}[htbp]
    \centering
    \includegraphics[width = 0.95 \textwidth]{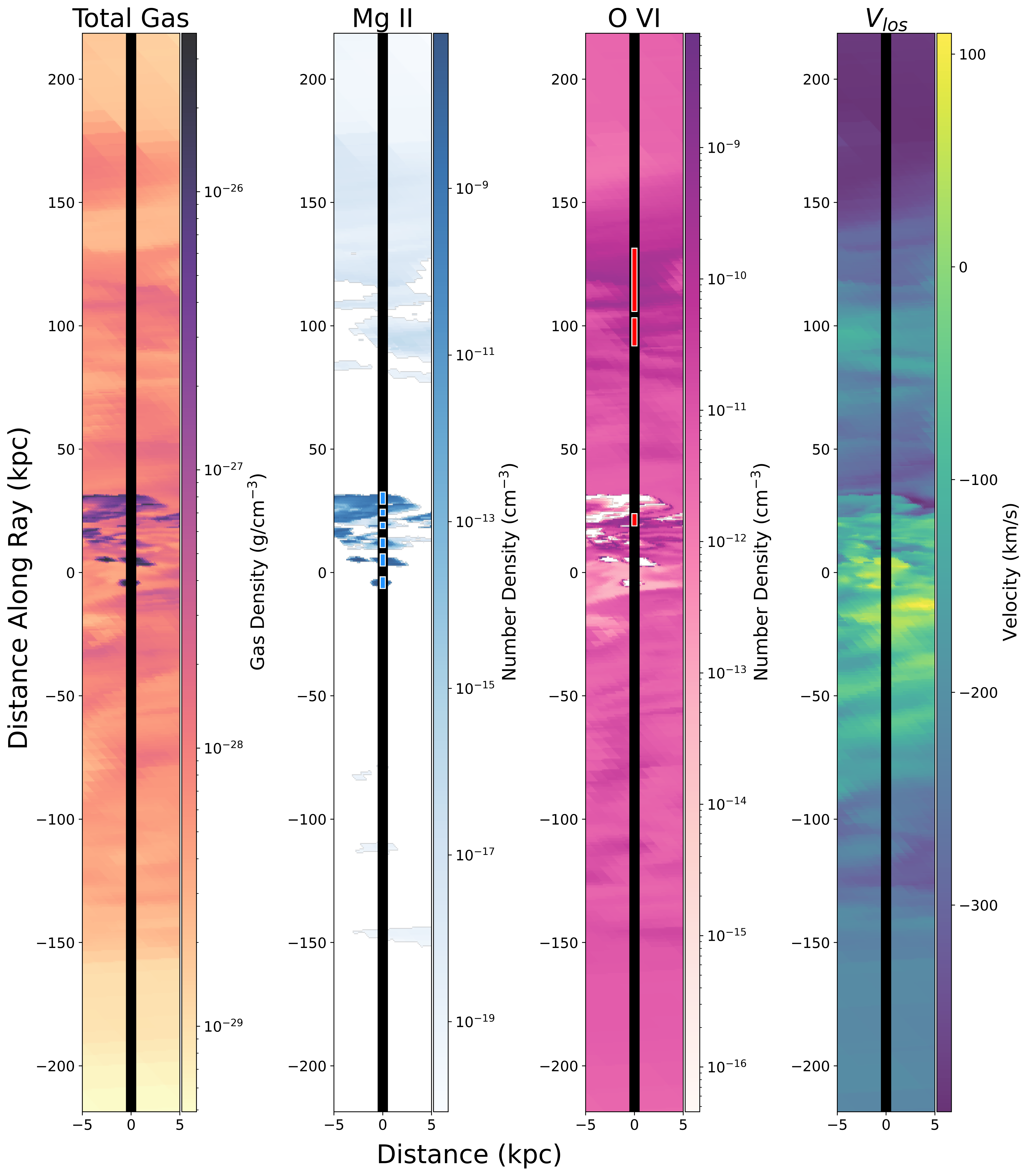}
    \caption{Two-dimensional slice plots through the FOGGIE galaxy Blizzard at $\mathrm{z=0}$ aligned with a one-dimensional SALSA ray that passes through the dataset (marked as the black vertical line through $\mathrm{x=0}$). 
    All four images from left to right show the same ray with slices in the total gas density, \mgii\ number density, \ovi\ number density and line of sight velocity ($\vec{v} \cdot \hat{r}$).
    A perpendicular distance of $\mathrm{\pm 5~kpc}$ is chosen arbitrarily to show the gas structures that the one-dimensional ray passes through. 
    The short blue and red lines drawn on top of the black ray line in the \mgii\ and \ovi\ number density plots represent locations of the mock absorbers detected by SALSA. 
    This ray, like all other rays in the dataset, passes through the entire virial radius, and only contains information from within the virial radius.  
    This ray passes through the galaxy at an impact parameter of 20.7~kpc.}
    \label{fig:ray_image}
\end{figure*}

Only a fraction of mock sightlines contain a column density above the \threshold\ threshold in each two dimensional dataset.
The probability of a mock sightline having \mgii\ or \ovi\ column density above the threshold is quantified in Figure \ref{fig: CoveringFractionCalculation}. 
This figure shows the cumulative distribution functions (CDF) over the column densities of both \mgii\ and \ovi\ where the black line is again our observable threshold. 
We can use these datasets to imitate covering fraction since these sightlines are uniformly distributed throughout the entire two dimensional virial radius.  
The observable covering fraction for \mgii\ column densities is around 5 -- 10\%, whereas the observable \ovi\ column densities covering fraction is closer to 65 -- 80\%. 
The average percentage above the threshold across the FOGGIE galaxies is reported for each redshift in each subplot of Figure \ref{fig: CoveringFractionCalculation}. 
The result is that a random sightline within the virial radius is much more likely to contain an observable column density of \ovi\ than \mgii.

There is a small but noticeable difference in the probability of \mgii\ column densities falling above the threshold as a function of redshift. 
At higher redshifts, we see a greater percentage of mock sightlines that have \mgii\ above the threshold. 
This is likely because the halos have smaller virial radii and cool gas extends farther out as the galaxy's disk is still forming. 

Figure \ref{fig: CoveringFractionCalculation} examines all mock sightlines within the virial radius equally; however, as we could deduce from the structure seen in Figure \ref{fig: Covering-Fraction-Profile}, these probabilities are dependent upon impact parameter. 
Figure \ref{fig:ObservableFraction_vs_ImpactParameter} shows the probability of mock sightlines containing column densities of 
just \mgii, just \ovi, or both above the observable threshold.
We can see at low impact parameter we have a high probability of a mock sightline containing both \mgii\ and \ovi, but around $\mathrm{(0.3 -0.5)R_{vir}}$  (depending on halo/redshift), the probability of observing both dips down to roughly zero.
The majority of mock sightlines begin to contain just \ovi\ column densities above the threshold and not \mgii\ column densities above the threshold. 
This result is consistent with trends seen in Figures \ref{fig: Covering-Fraction} and \ref{fig: Covering-Fraction-Profile} where there are few, if any, \mgii\ column densities above the threshold at larger impact parameters. At large impact parameter all of the lines move towards zero, meaning that there are many sightlines that do not contain observable column densities of either \mgii\ or \ovi.

The key takeaway from of Figure \ref{fig:ObservableFraction_vs_ImpactParameter} is that for low impact parameters ($\mathrm{\lesssim 0.3 R_{vir}}$) we would expect to observe both \mgii\ and \ovi\ along a line of sight, and at mid to high impact parameters, we would only expect to observe \ovi.
We would almost never expect to have a column density of \mgii\ above the threshold without there also being a column density of \ovi\ above the threshold. 
This is true across all halos and redshifts.

\section{Mock Absorption Features along Sightlines}\label{sec:SALSA}
We investigate the physical structures of gas along a sightline so we can compare how specific clumps or overdensities of gas interact with each other. 
We identify these overdensities using the software package SALSA, Synthetic Absorption Line Surveyor Application \citep{boyd_salsa_2020}. 
SALSA identifies physically and kinematically contiguous overdensities of individual ion column densities along a one dimensional ray through the simulation.
All of these absorbers achieve a column density greater than \threshold.
It compiles these overdensities into a set of mock absorbers for that ray (specifics on how SALSA identifies mock absorbers can be found in Appendix \ref{sec: app}). 
For this paper, we use the SPICE algorithm in the SALSA package to identify mock absorbers.   

At each redshift for each halo we construct 400 one-dimensional rays with SALSA at random orientations and with impact parameters between 5--40\% the virial radius. We extract the \mgii\ and \ovi\ mock absorbers from these rays. Figure \ref{fig:ray_image} gives a visual representation of just one of these rays in gas density, \mgii\ density, \ovi\ density, and line of sight velocity. The black line in these images is the one-dimensional trajectory of the SALSA ray and a slice, or two-dimensional plane, is rendered around the ray to demonstrate the properties of the plasma that the ray is passing through. The red and blue lines on top of the black lines are the \mgii\ and \ovi\ absorbers that SALSA picks out for this ray. We use this figure as an intuition-building example to guide the reader's understanding for how the dataset used in this section is constructed.  

Section \ref{subsec: absrobers v impact parameter} will discuss the probability of a ray containing just \ovi\ mock absorbers, just \mgii\ mock absorbers, or both versus impact parameter. Section \ref{subsec: Distance Velocity} will discuss the difference and distance and line of sight velocity of pairs of \mgii\ and \ovi\ mock absorbers.

\subsection{Absorber Detection Probabilities}\label{subsec: absrobers v impact parameter}

\begin{figure*}[htbp]
    \centering
    \includegraphics[width=\linewidth]{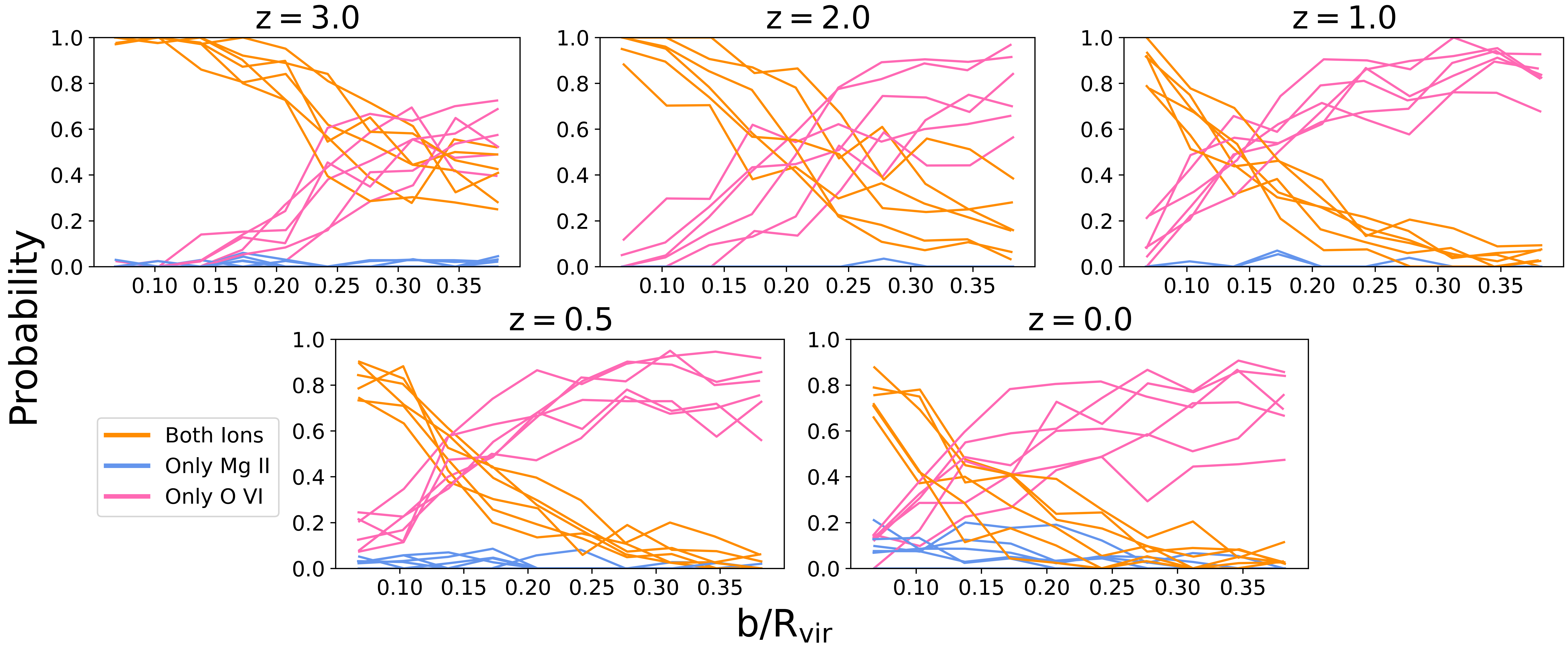}
    \caption{The probability of detecting just \mgii\ mock absorbers, just \ovi\ mock absorbers, or both mock absorbers within a ray as a function of impact parameter.   The orange represents the probability of a ray having both mock absorbers, the blue represents the probability a ray having only \mgii\ mock absorbers and the pink represents the probability of a ray having only \ovi\ mock absorbers. Each line for each color represents one of the six FOGGIE halos. All mock absorbers identified by SALSA are inherently above our threshold of \threshold. We report these distributions at redshifts of $\mathrm{z = 3,~2,~1,~0.5,~and~0}$.}
    \label{fig:salsa_absrober_fractions}
\end{figure*}

Similar to Figure \ref{fig:ObservableFraction_vs_ImpactParameter}, we can look at the ``observable probability" vs. impact parameter.
However, the method and meaning behind this calculation in Figure \ref{fig:salsa_absrober_fractions} is different than Figure \ref{fig:ObservableFraction_vs_ImpactParameter}.
Before, with the two-dimensional mock sightline images the threshold of \threshold\ applied to the total integrated column density over the whole ray. 
SALSA works differently, identifying physically distinct over-dense ``clumps'' of specific ion along the line of sight that would likely make up a single and distinct absorption feature. 
Each one-dimensional clump must have a total column density greater than \threshold\ to be identified. 
Both methods produce similar trends on a broad scale, but we discuss the differences between these two methods and what this indicates about the results in Section \ref{sec: model-diffs}. 
Here we focus on just the results from the SALSA analysis itself. 

We see in Figure \ref{fig:salsa_absrober_fractions} that at low impact parameter we have a high probability of both \mgii\ and \ovi\ absorbers being identified in the ray. 
This gradually drops off with impact parameter and at large impact parameters we expect to see mostly \ovi\ absorbers alone in the rays. 
This is largely fueled by the central nature of the \mgii-bearing gas, as previously discussed. 
There is a slightly greater chance of picking out \mgii\ absorbers in a ray without picking up \ovi\ absorbers at low impact parameter than there was in Section \ref{sec: 2D results}, particularly at low redshift.

There is also some redshift dependence. 
We can see that as redshift decreases the probability of observing both \mgii\ and \ovi\ absorbers reaches nearly zero at a lower impact parameter than it did with the analysis in Section \ref{sec: 2D results}.
At $\mathrm{z = 0}$ we reach approximately zero probability of observing both ions around $\mathrm{(0.15-0.25) R_{vir}}$, whereas in Figure \ref{fig:ObservableFraction_vs_ImpactParameter} we reach this over a range of $\mathrm{(0.2-0.4) R_{vir}}$.
Overall, this trend is due to the \mgii-bearing gas becoming more and more centrally located as the galaxy settles into its isolated state.

\subsection{Spatial and kinematic relationships between absorbers} \label{subsec: Distance Velocity}

\begin{figure*}[htbp]
    \centering
    \includegraphics[width=\linewidth]{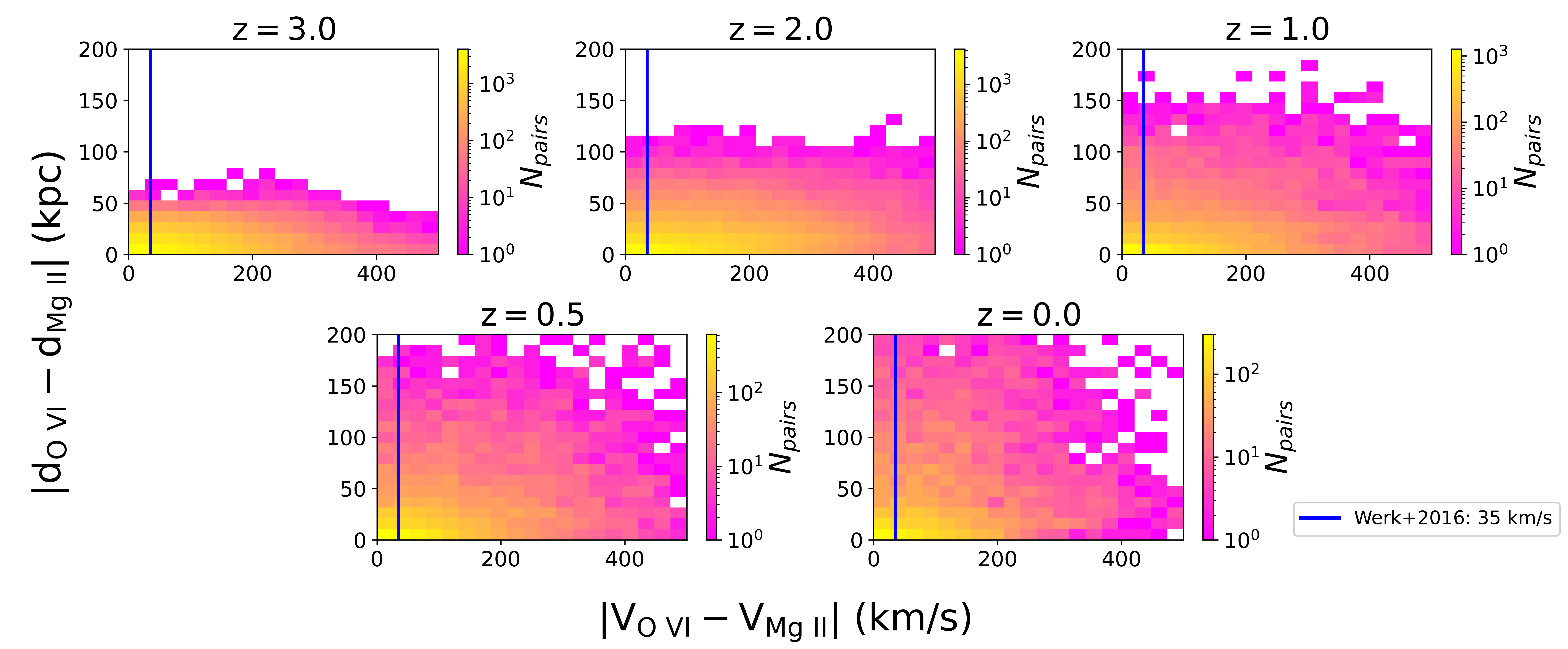}
    \caption{Two-dimensional histogram of the distance between pairs of \mgii\ and \ovi\ mock absorbers and the column density weighted average line of sight velocity of these mock absorber pairs. 
    All pairs are limited to the same ray, but if a single ray contains multiple \mgii\ and \ovi\ mock absorbers all pairs of absorbers are considered. 
    The color represents the number of absorber pairs that fall within the binned range.
    The blue vertical line represents 80\% of O VI absorbers being within 35 km/s of a low-ion absorber from  \citet{werk_cos-halos_2016}. Each panel contains mock absorber pairs from all six FOGGIE halos. We report these distributions at redshifts of $\mathrm{z = 3,~2,~1,~0.5,~and~0}$. Note the distance difference is limited to $\mathrm{2R_{vir}}$, and since virial radii are smaller at larger redshift we see a distance cutoff in the high redshift distributions. }
    \label{fig: DV}
\end{figure*}

We can explore the kinematic relationship between pairs of mock absorbers by looking at difference in line of sight velocity that occurs between \mgii\ and \ovi\ mock absorbers. 
Figure \ref{fig: DV} shows the relationship between the difference in column density weighted average line of sight velocity and the difference in line of sight distance between \mgii\ and \ovi\ absorber pairs. 
Absorber pairs are limited to \mgii\ and \ovi\ mock absorbers along the same ray since we aim to model QSO sightlines through the CGM. 
However, if multiple \mgii\ and \ovi\ absorbers occur along a ray, all combinations of \mgii\ and \ovi\ pairs are considered. 
The number of absorber pairs that occur along our mock sightlines vary, but we keep consistent the total number of rays at each redshift for each halo at 400. 
All of the halos are combined to make each subplot in Figure \ref{fig: DV} as the structure and distribution of these ions across all of the previous figures show that they are in remarkable agreement. 
SALSA keeps track of the physical extent of each mock absorber along the ray and we compare the geometric midpoints (average of the of start and end points) of these mock absorbers to get the relative distance between mock absorber pairs.

We can see in Figure \ref{fig: DV} that there are a wide range in velocity differences and distance differences between the mock absorber pairs. 
It is important to note that the distance at higher redshift cannot extend as high as lower redshift since we limit that data to within the virial radius, and the virial radii are smaller at high redshift. The most populated bins of the two-dimensional histogram have low distance difference and low velocity. However, the data are fairly spread out across 100s of km/s in velocity space and 10s to 100s of kpc in distance space. Many velocity separations lay above the \citet{werk_cos-halos_2016} results shown in blue.
Even some of the pairs that are within  $\mathrm{35~km/s}$ are not necessarily spatially correlated with one another.
Overall, this would suggest that these two ions, while having some correlation, are not as correlated in the FOGGIE simulations as we see in the COS-Halos QSO absorption line observations of \citet{werk_cos-halos_2016}. 
This is discussed in more depth in Section \ref{subsec: absrober relationship}.

\section{Discussion}\label{sec: discussion}

This section furthers the discussion of our results, comparing to observations, considering the differences between analysis methods, and examining the caveats of our analysis. 
Section \ref{subsec: column-density-comparison} compares the bulk distribution we see for \mgii\ and \ovi\ column densities to observations. 
Section \ref{subsec: absrober relationship} discusses the spatial and kinematic relationship between \mgii\ and \ovi\ absorbers, connecting our results to relevant observations. 
Section \ref{sec: model-diffs} examines the additional information we can gain by comparing our analyses from Section \ref{sec: 2D results} and Section \ref{sec:SALSA}. 
Finally, Section \ref{subsec: caveats} outlines some of the caveats of the results and analysis of this paper. 

\subsection{Distribution of \mgii\ and \ovi-bearing gas} \label{subsec: column-density-comparison}

It is clear from the results in Figures \ref{fig: Covering-Fraction-Profile}, \ref{fig: CoveringFractionCalculation}, and \ref{fig:ObservableFraction_vs_ImpactParameter} that observable amounts of \mgii\ column density (greater than \threshold) occupy only a small covering fraction inside of the virial radius. 
\mgii-bearing gas exists mostly in the central regions around the galaxy, whereas the \ovi-bearing gas exists as a significant, space-filling diffuse halo out to larger impact parameters. 
This is consistent with previous observations of gas distribution in Milky Way-like galaxies from population studies. 
Figure 2 in \citet{tumlinson_circumgalactic_2017} shows this radial profile for many different ions across multiple population studies and \citet{lehner_project_2025} shows a detailed example for a single galaxy, Andromeda. 
We see very clearly that low-ions (i.e., ions tracing cool gas) drop off more steeply in impact parameter than high-ions (i.e., ions tracing gas with temperatures closer to the virial temperature).

This diffuse cloud of \ovi-bearing gas would yield broader line widths due to the Doppler broadening given
\begin{equation}
    \lambda_{obs} = \lambda_0 \left( 1+ \frac{v_{gas}}{c} \right) \label{equ: DB}
\end{equation}
If we assume a rest frame wavelength of $\sim\mathrm{1000 \AA}$ (appropriate for \ovi) and a maximum velocity spread of roughly $\mathrm{v_{gas} \sim 10^3~km/s}$ (typical for a Milky-Way like galaxy) we would expect a broadening of a few $\mathrm{\AA}$ due to this halo structure. 
We do see in observations of \ovi\ that it tends to have a more broad profile \citep{werk_cos-halos_2016, qu_cosmic_2024}.

\subsection{Relationship between \mgii\ and \ovi\ absorbers} \label{subsec: absrober relationship}

\begin{figure*}
    \centering
    \includegraphics[width=\linewidth]{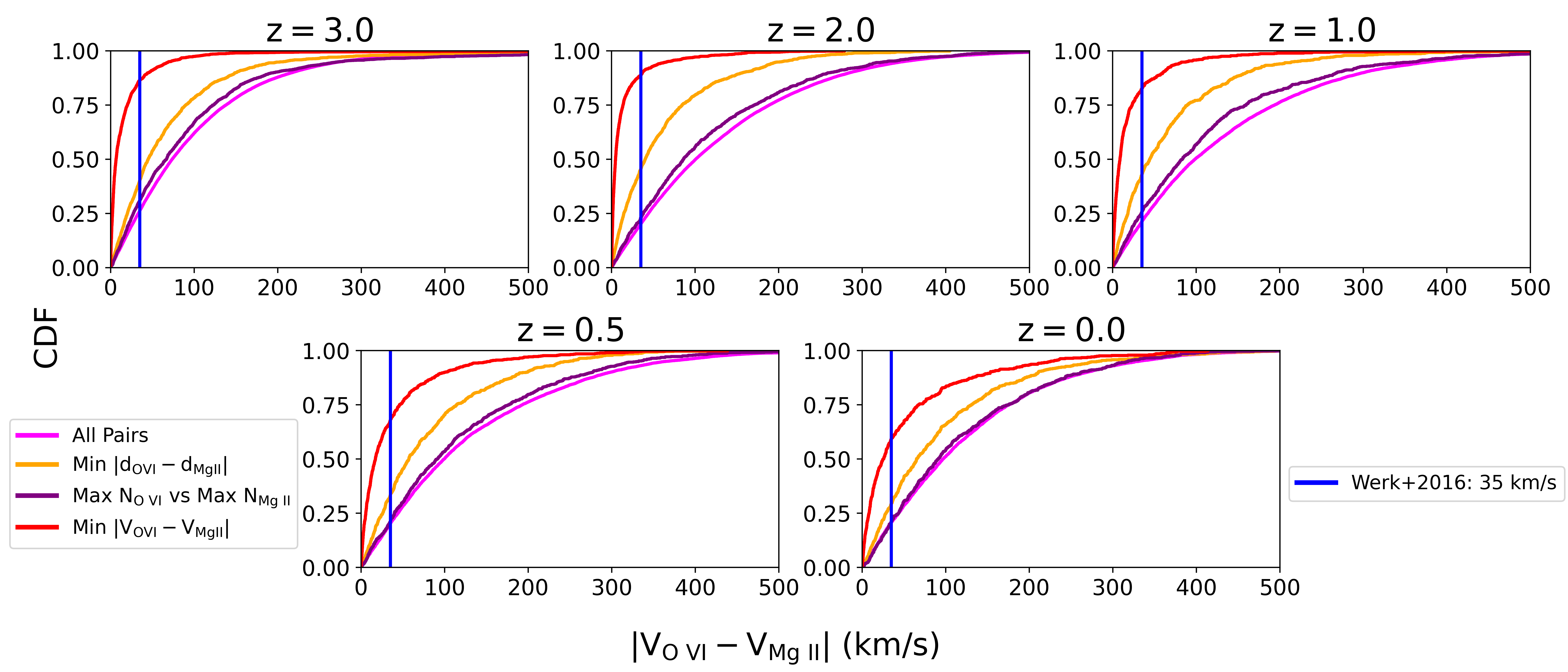}
    \caption{Cumulative distribution functions (CDF) of velocity difference between \mgii\ and \ovi\ mock absorber pairs. 
    The blue vertical line represents the 80\% of \ovi\ absorbers being within 35 km/s of a low-ion absorber.  
    We consider 4 different ``cutoffs" for our CDFs: we look at the velocity difference between all pairs of \mgii\ and \ovi\ mock absorbers (pink), only the closest pairs within a ray (orange), the single largest column density pair of \mgii\ and \ovi\ mock absorbers along the sightline (purple), and the mock absorbers pair with the smallest velocity separation between them (red).
    We report these distributions at redshifts of $\mathrm{z = 3,~2,~1,~0.5,~and~0}$.}
    \label{fig: velocity_CDF}
\end{figure*}

There is a kinematic correspondence between \ovi\ and low-ions in observations even though based on the density and temperature phases these ions exist at we would not naively expect this \citep[e.g.,][]{werk_cos-halos_2016, tripp_highresolution_2008}. 
 These observations implicitly suggest a connection between high-ion and low-ion absorbers that could illuminate the structure of the circumgalactic gas. 
In Figure \ref{fig: velocity_CDF} we probe this relationship with four different cumulative distribution functions examining the cokinematic structure of \ovi\ and \mgii\ mock absorbers in our SALSA sightlines. 
The pink line considers all \mgii\ and \ovi\ pairs at or above the threshold column density.  
The orange line considers only the physically closest absorbers along each ray. 
The purple line represents only the single largest column density pairing along each ray and the red line represents the absorber pair with the smallest velocity difference along the line of sight.   
The blue vertical line is at $\mathrm{35~km/s}$, which comes from  \citet{werk_cos-halos_2016} where approximately 80\% of \ovi\ absorbers had a low-ion absorber within $\mathrm{35~km/s}$. 
The red line is broadly consistent with the majority of the \ovi\ mock absorbers having a \mgii\ mock absorber within $\sim \mathrm{35~km/s}$, and is most consistent with the comparison methodology used in \citet{werk_cos-halos_2016}.
However, the physically closest \mgii\ and \ovi\ absorbers (as shown in orange in Figure \ref{fig: velocity_CDF}) tend to have much larger velocity separations. 
This continues to be true for the largest column density pairing and all of the pairings in the sightlines.
This indicates that high velocity coincidence of between single \ovi\ and \mgii\ absorbers along sightlines does not necessarily mean high spatial coincidence of these absorbers.
Since most rays have multiple mock absorbers of each ion, we are seeing a selection effect where absorbers that are in no way physically related appear to be cokinematic. 
This may also be true for observations of these ions and seems to more closely align with the intuition that these ions trace different thermodynamic phases of gas and therefore should not generally be co-=spatial. 
We note the same selection effect as shown in Figure \ref{fig: velocity_CDF} would occur in distance separation space. The line of least distance separation would rise the most quickly if you plotted against the distance separation between mock absorber pairs.

\begin{figure*}
    \centering
    \includegraphics[width=1\linewidth]{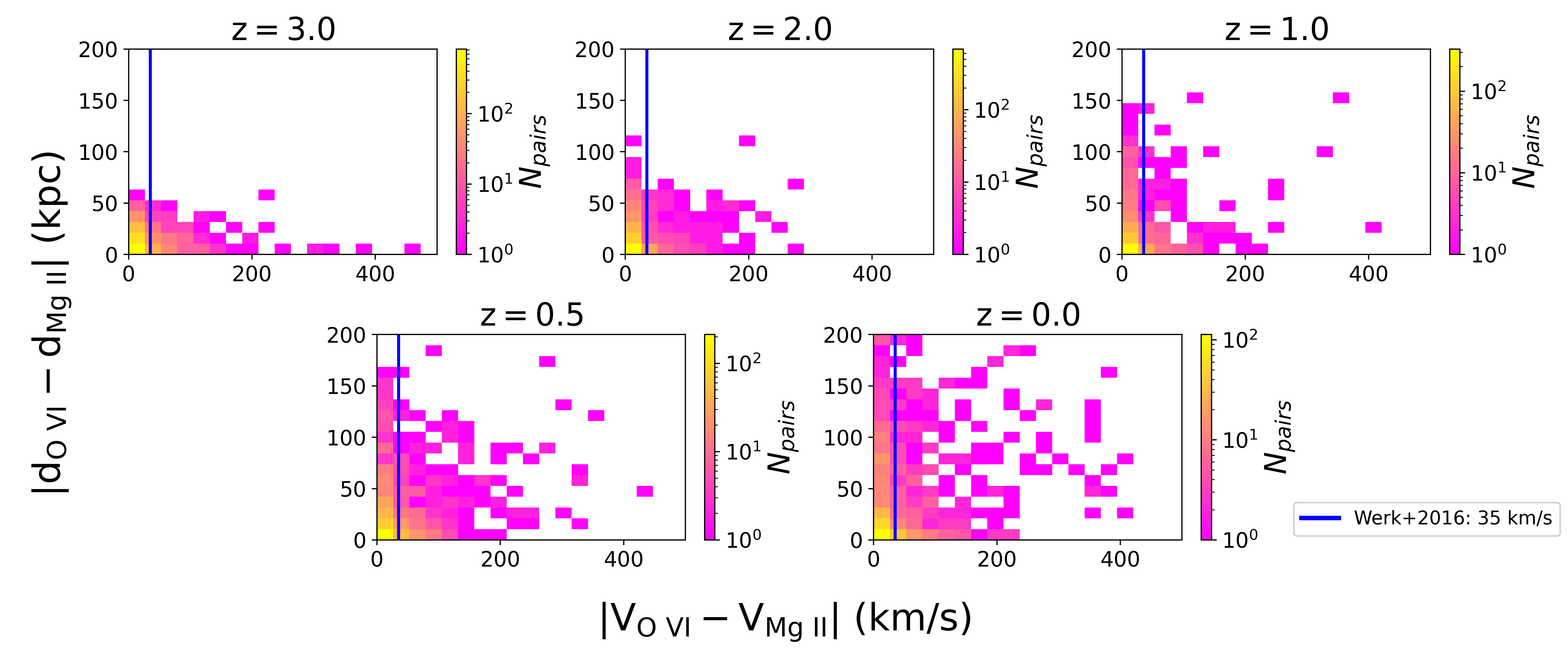}
    \caption{Two-dimensional histogram akin to Figure \ref{fig: DV}. 
    We show the distance between pairs of \mgii\ and \ovi\ mock absorbers and the column density weighted average line of sight velocity for the pairs of mock absorbers that have the smallest velocity separation along the sightline. 
    The color represents the number of absorber pairs that fall within the binned range. 
    The blue vertical line represents 80\% of O VI absorbers being within 35 km/s of a low-ion absorber from  \citet{werk_cos-halos_2016}. 
    Each panel contains mock absorber pairs from all six FOGGIE halos. We report these distributions at redshifts of $\mathrm{z = 3,~2,~1,~0.5,~and~0}$. 
    Note the distance difference is limited to $\mathrm{2R_{vir}}$, and since virial radii are smaller at larger redshift we see a distance cutoff in the high redshift distributions.}
    \label{fig:DV_closest_v}
\end{figure*}

Figure \ref{fig: velocity_CDF} hints at a breakdown of the correspondence between the cokinematic and cospatial nature of these mock absorber pairs. 
We can see this point even more clearly in Figure \ref{fig:DV_closest_v}, which shows the same thing as Figure \ref{fig: DV}, but now only for the absorbers with the closest velocity separation along the sightline (our red line from Figure \ref{fig: velocity_CDF}). 
It is clear from Figure \ref{fig:DV_closest_v} that the absorber pairs that have the smallest velocity separation are not necessarily cospatial with many pairs being tens to hundreds of kpc away from one another. 
This highlights the lack of a clear spatial relationship between absorbers that are cokinematic.

Lastly, we note that the \citet{werk_cos-halos_2016} paper differentiates between ``broad" (high-velocity dispersion) and ``narrow" (low-velocity dispersion) \ovi\ absorbers. However, in \citet{werk_cos-halos_2016} the number of broad and narrow \ovi\ absorbers that correspond to low-ion absorbers are approximately the same. Therefore, we do not split up our sample of mock  \ovi\ absorbers into broad and narrow categories, since the absorber width does not seem to have a large effect. However, we note that our sample of mock \ovi\ absorbers has a smaller fraction of broad absorbers than is found in \citet{werk_cos-halos_2016}. This discrepancy is likely due to the different selection methods of SALSA mock absorbers, which depend on both spatial and kinematic information, versus observational absorbers, which are selected solely based on kinematic observations. More information about the SALSA method can be found in Appendix \ref{sec: app}.

\subsection{Differences between Integrated Column Density Analysis and Absorption Feature Analysis}\label{sec: model-diffs}

\begin{figure}[hbtp]
    \centering
    \includegraphics[width=\linewidth]{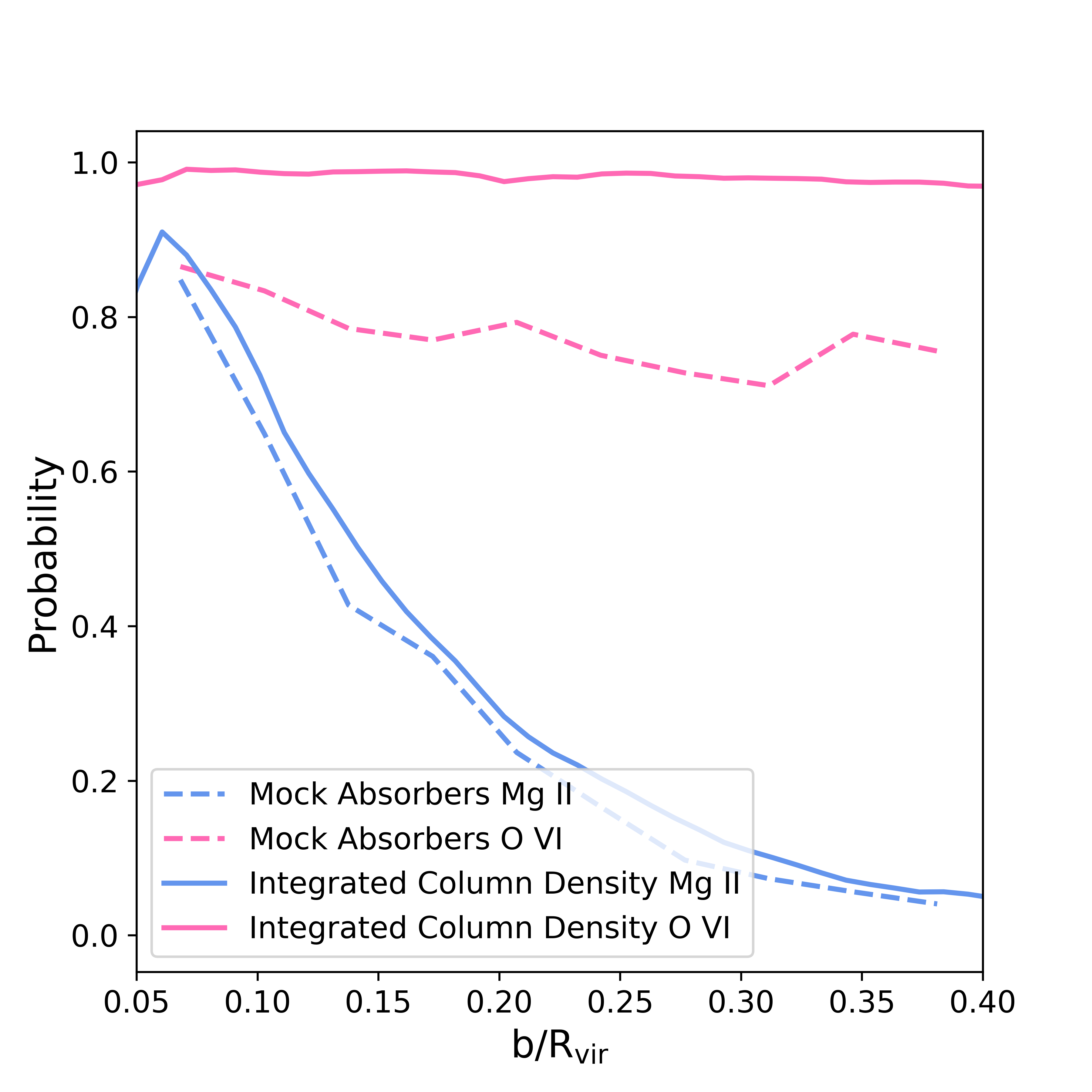}
    \caption{Comparison of the probability of detection for our two different analysis methods, the two-dimensional mock sightlines grids and the the one-dimensional SALSA mock absorber analysis.
    The solid lines above represent the two-dimensional integrated mock sightline analysis and the dashed lines represent the SALSA mock absorber analysis.    The probability of the two-dimensional datasets represents the probability of the total integrated column density being larger than \threshold\ whereas the probability for the SALSA rays represents whether SALSA finds a spatially and kinematically contiguous absorbers above \threshold\ along the sightline.
    The blue lines represent the probability of detecting a \mgii\ structure for each analysis and the pink lines are the probability of detecting a \ovi\ structure for each analysis. 
    Each line is an average across all of the FOGGIE halos at $\mathrm{z=0}$. 
    }
    \label{fig:2D-Salsa_Comparisons}
\end{figure}

Both of the methods discussed in Sections \ref{sec: 2D results} and \ref{sec:SALSA} are typical ways that one can analyze simulations to more easily compare them to observations.
The mock sightlines in Section \ref{sec: 2D results} are integrated over the entire line of sight and are identified as ``observable" if the column density exceeds \threshold. 
The mock sightlines in Section \ref{sec:SALSA} are characterized only by individual absorbers along the line of sight, where absorbers must be physically continuous and have a smooth velocity gradient.  
Looking at the results of these two methods together, we can deduce the properties that diffuse gas plays in column density measurements. 
This diffuse gas contributes to the integrated lines of sight, but would not be identified as part of a physical absorber in Section \ref{sec:SALSA}.
We can see a comparison of these two methods in Figure \ref{fig:2D-Salsa_Comparisons}. 
This figure compares two different probabilities, the probability of a sightline having a mock absorber or the probability of a sightline having a total integrated column density above the threshold. Probability is shown as a function of impact parameter. 

We can see in Figure \ref{fig:2D-Salsa_Comparisons} that the average probability of a mock sightline containing an \ovi\ absorber is smaller than the probability of the total integrated column density exceeding the threshold at all impact parameters. 
This suggests that the smaller, cumulative amount of \ovi-bearing gas makes significant contributions when integrating along the entire line of sight. 
This makes observations complicated as diffuse gas throughout the CGM creates a wide absorption feature due to Doppler broadening as discussed in Section \ref{subsec: column-density-comparison}. 
Therefore, we expect that observations of \ovi\ lines would be broad and we do see observational evidence for this in \ovi\ spectral features \citep[in, e.g.,][]{werk_cos-halos_2016}. 

The effects of diffuse \mgii\ are much smaller as the probability distributions in Figure \ref{fig:2D-Salsa_Comparisons} are much closer. 
Therefore, we expect only a small fraction of sightlines to have diffuse \mgii\ that would give an observable integrated column density without producing an absorber. 
This means in a population of mock sightlines \mgii-bearing gas is more effectively modeled as absorption features, with only smaller contributions from diffuse components.

\subsection{Caveats} \label{subsec: caveats}

The analyses in this paper are subject to a variety of challenges and limitations. These include: 

\begin{enumerate}
        \item The simulations themselves implement a specific mode of galaxy formation and the choice of physics for these simulations can affect the structure of the CGM. The FOGGIE simulations do not necessarily implement all possible physics and lack prescriptions for magnetic fields and cosmic rays, which may affect the structure of the CGM for these simulated galaxies \citep{vandevoort_effect_2021,weber_crexit_2025, butsky_impact_2022}.
        This is also true for our choice of physics in post-processing; for example, the UV background \citep[we use][]{haardt_radiative_2012} has been shown to greatly affect the resulting ionic column densities in post processing \citep{taira_impacts_2025}. 
        The effects of these physics choices are studied by the AGORA Collaboration, and a synthesis of these effects on the CGM can be found in \citet{strawn_agora_2024}.
        Since this paper heavily relies on these ionic column densities, it is important to note that this is a model-dependent quantity. 
        \item Additionally, we are working with a limited sample size of six galaxy halos and in a very specific mass regime ($\mathrm{M_{vir} \approx (0.50- 1.6) \times 10^{12}~M_\odot}$ at $\mathrm{z=0}$). 
        These factors both affect the scalability of our study and whether these trends should apply to galaxies more widely. 
        \item We define a set ion column density threshold of \threshold\ in Section \ref{sec: 2D results} to denote observable \mgii\ and \ovi\ column densities. 
        This is only an approximation of the observational column density threshold for these ions, and real observations have numerous factors that affect observability, including observation duration, spectral resolution, signal-to-noise, etc.
        Additionally, this value of \threshold\ is a discretionary choice made by the authors based on current observational capabilities. 
        Changing this value would have small changes on particularly the number of sightlines with observable \ovi\ and the number of \ovi\ mock absorbers, since \ovi\ column densities tend to be closer to the threshold. However, any reasonable changes in $\mathrm{N}$ only create small quantitative differences and the overall qualitative behavior and conclusions in this paper remain the same.
        \item There are multiple parameters that go into the SALSA algorithm that identify individual absorption features that we see in Section \ref{sec:SALSA}. 
        We outline in Appendix \ref{sec: app} what these parameters are and the values we use for this analysis. 
        We note that reasonable changes in these parameters would not have a significant effect on this analysis.
        \item We assume ionization equilibrium when extracting ion densities in our simulation. This overlooks any effects of nonequilibrium contributions to the ionization fractions, such as the time delay in \ovi\ to \ion{O}{5} recombination.
        \item We have compared two ways of estimating column densities of individual ions --- projected column densities through the entire CGM and an algorithm that identifies individual physically-contiguous structures along a line of sight. 
        Both are approximations of actual astronomical observations in the sense that they lack direct information about the detailed absorption features that may emerge (e.g., line blending from physically distinct features that absorb light at similar wavelengths). 
        More work on the relationship between mock observations and real observations can be found in \citet{marra_using_2021, marra_examining_2023}.
        A more realistic method of identifying and relating low- and high-ion gas would be to create and analyze synthetic absorption spectra and analyze them as if they were real astronomical spectra (which is possible using the Trident code). 
        We have chosen not to do so for this project, but rather use a method that focuses on physical structures in galactic systems. 
        We have done our analysis this way as there is uncertainty in tying spectral features to underlying gas and this work was focused on looking at the underlying gas structures. 
\end{enumerate}


\section{Summary}\label{sec: summary}

We have used the FOGGIE simulations to probe the relationship between \mgii\ and \ovi\ as tracers for low and high-ions, or cool and warm gas, respectively. These Milky Way-like galaxy simulations are constructed to have high  spatial resolution far out into the simulations' circumgalactic medium, which allows us to effectively probe gas features out to significant fractions of the virial radius.

The key results from this work are as follows:

\begin{enumerate}
    \item There is an extended halo of \ovi\ that yields column densities greater than \threshold\ out to impact parameters that are a significant fraction of the virial radius.  In contrast, \mgii\ has a radial profile that rapidly falls below this column density threshold, meaning that most detectable amounts of \mgii\ are located close to the galactic disk in all systems that we examined and over a wide range of observable redshifts.
    \item Two-dimensional projected mock sightlines show that for low impact parameters we would expect to have both \mgii\ and \ovi\ at column densities exceeding \threshold\ in our lines of sight, but above roughly $\mathrm{0.3 - 0.5 R_{vir}}$ we expect this to change to sightlines having only detectable amount of \ovi. 
    \item The SALSA one-dimensional ray analysis finds individual physically contiguous structures in a given ion along the mock line of sight. This analysis shows notably lower rates of \ovi\ detection than the two-dimensional projected sightline analysis. This reinforces the idea that \ovi-bearing gas exists as a diffuse and extended halo around the galaxy and is not easily classified into single physically over-dense features.
    \item We find that \mgii\ and \ovi\ absorbers appear to be kinematically correlated  only when looking at the absorbers with the closest line of sight velocities along a sightline. However, this does not correspond to any physical spatial correlation of these mock absorbers and therefore there is no actual co-movement of the gas that traces these ions. 
    \item Overall the trends in this work are strikingly consistent across redshift and across simulated galaxies. There is some scatter across the six FOGGIE halos, but most plots have remarkable agreement for the six different galaxies. 
    Evolution in redshift is very slight and does not significantly affect any of the trends presented in this paper.  

\end{enumerate}

Overall, more work needs to be done to understand the relationship between the cooler gas that is found at small fractions of the virial radius and the diffuse warm medium that surrounds it. Probing the relationship between these states and the interfaces between the cool and warm gas of the circumgalactic medium is key to broadening our understanding of gas dynamics in galaxies.

\begin{acknowledgments}

MT acknowledges support from Michigan State University through the University Distinguished Fellowship program and from the National Science Foundation through the Graduate Research Fellowship Program grant number \#2235783.
BWO acknowledges support from NSF grants \#1908109 and \#2106575,
NASA ATP grants 80NSSC18K1105 and 80NSSC24K0772, and NASA TCAN grant 80NSSC21K1053.
CL was supported by NASA through the NASA Hubble Fellowship grant \#HST-HF2-51538.001-A awarded by the Space Telescope Science Institute, which is operated by the Association of Universities for Research in Astronomy, Inc., for NASA, under contract NAS5-26555.
CWT was supported for this work in part by NASA via a
Theoretical and Computational Astrophysics Networks
grant \#80NSSC21K1053 and JWST AR \#5486.
RA acknowledges funding from the European Research Council (ERC) under the European Union's Horizon 2020 research and innovation programme (grant agreement 101020943, SPECMAP-CGM).

Computational resources supporting this work were provided by the NASA High-End Computing (HEC) Program through the NASA Advanced Supercomputing (NAS) Division at Ames Research Center and were sponsored by NASA's Science Mission Directorate; we are grateful for the superb user-support provided by NAS. Resources were also provided by the Blue Waters sustained-petascale computing project, which is supported by the NSF (award numbers ACI-1238993 and ACI-1514580) and the state of Illinois. 
This work also used the resources of the Michigan State University High Performance Computing Center, operated by the MSU Institute for Cyber-Enabled Research. Simulations described in this work were performed using the publicly-available Enzo code, which is the product of a collaborative effort of many independent scientists from numerous institutions around the world. 

\end{acknowledgments}


M.T. contributed to the conceptualization, investigation,
formal analysis, visualization, and writing (original draft) of
this paper. 
B.W.O. contributed to the conception, supervision, and writing (review and editing) of this paper.
C.K. contributed to the software, methodology, and writing (review and editing) of this paper. 
C.L., M.P., J.T., C.T., R.A., N.L., B.D.S., and J.C.H. contributed to the conceptualization and writing (review and editing) of this paper. 
V.S. contributed to the conceptualization of this paper.


\software{Enzo \citep{bryan_enzo_2014,brummel-smith_enzo_2019}, 
yt \citep{Turk2011}, Trident~\citep{TridentMethods}, SALSA~\citep{boyd_salsa_2020}, {\sc matplotlib} \citep{hunter2007}, {\sc numpy} \citep{walt2011numpy}, \textsc{scipy} \citep{scipy2020}, astropy \citep{Astropy1,Astropy2}, ChatGPT \citep{chatgpt}, Claude \citep{claude_opus}.}

\section{Appendix}
\section{Absorption identification methods in SALSA} \label{sec: app}

SALSA \citep{boyd_salsa_2020} is designed to identify overdensities of gas along a single 1D line of sight through a simulation. 
These overdensities of gas are physically and kinematically contiguous.
The main algorithm behind SALSA works to identify which overdense regions, which we term mock absorbers, take up the majority of the total integrated column density in the sightline.
Therefore our first step is to find the total ion column density along a line of sight.
Then it sets a number density threshold such that 80\% of the column density is contained in cells above that threshold. Then SALSA identifies which the distinct regions above this threshold and classifies these as mock absorbers.
Next, SALSA masks out the regions that have already been identified and again sets a line at 80\% the total column density of the remaining gas. 
Then it checks if any newly identified absorbers can be combined with absorbers from the previous steps by checking if they connect spatially and the average line of sight velocities have a difference of less than $\mathrm{10~km/s}$. 
If not, these new absorbers are added to the list of mock absorbers for that ray.
It iteratively performs these steps again until the total column density of the species in question that is not contained within an absorber drops below \threshold\ (which is typically a very small fraction of the total line of sight column density).
Finally, SALSA checks to see if all of the absorbers identified reach the minimum observable value, which is the same threshold that terminates the iteration process --- \threshold.
More information about the SALSA algorithms can be found on the documentation website \footnote{\url{https://salsa.readthedocs.io/en/latest/absorber_extraction.html##spice-method}}

The 80\% column density threshold, the velocity threshold ($\mathrm{10~km/s}$), and the minimum column density threshold \threshold\ are all parameters that are set by the algorithm and can be changed by the user. 
We choose to use the default parameters for the 80\% column density threshold and the velocity threshold. We change the minimum column density from its default value of $\mathrm{N = 10^{13}~cm^{-2}}$ to \threshold\ to match the other analysis in the paper.

\bibliography{bibliography}{}

@ARTICLE{TridentMethods,
       author = {{Hummels}, Cameron B. and {Smith}, Britton D. and {Silvia}, Devin W.},
        title = "{Trident: A Universal Tool for Generating Synthetic Absorption Spectra from Astrophysical Simulations}",
      journal = {\apj},
         year = 2017,
        month = sep,
       volume = {847},
       number = {1},
          eid = {59},
        pages = {59},
          doi = {10.3847/1538-4357/aa7e2d},
archivePrefix = {arXiv},
       eprint = {1612.03935},
 primaryClass = {astro-ph.IM},
       adsurl = {https://ui.adsabs.harvard.edu/abs/2017ApJ...847...59H}
}

@ARTICLE{2013CLOUDY,
       author = {{Ferland}, G.~J. and {Porter}, R.~L. and {van Hoof}, P.~A.~M. and {Williams}, R.~J.~R. and {Abel}, N.~P. and {Lykins}, M.~L. and {Shaw}, G. and {Henney}, W.~J. and {Stancil}, P.~C.},
        title = "{The 2013 Release of Cloudy}",
      journal = {\rmxaa},
         year = 2013,
        month = apr,
       volume = {49},
        pages = {137-163},
          doi = {10.48550/arXiv.1302.4485},
archivePrefix = {arXiv},
       eprint = {1302.4485},
 primaryClass = {astro-ph.GA},
       adsurl = {https://ui.adsabs.harvard.edu/abs/2013RMxAA..49..137F}
}

@article{peeples_figuring_2019,
    title = {Figuring {Out} {Gas} \& {Galaxies} in {Enzo} ({FOGGIE}). {I}. {Resolving} {Simulated} {Circumgalactic} {Absorption} at 2 ≤ z ≤ 2.5},
    volume = {873},
    issn = {0004-637X, 1538-4357},
    url = {https://iopscience.iop.org/article/10.3847/1538-4357/ab0654},
    doi = {10.3847/1538-4357/ab0654},
    language = {en},
    number = {2},
    urldate = {2025-05-28},
    journal = {The Astrophysical Journal},
    author = {Peeples, Molly S. and Corlies, Lauren and Tumlinson, Jason and O’Shea, Brian W. and Lehner, Nicolas and O’Meara, John M. and Howk, J. Christopher and Earl, Nicholas and Smith, Britton D. and Wise, John H. and Hummels, Cameron B.},
    month = mar,
    year = {2019},
    pages = {129},
}

@article{corlies_figuring_2020,
    title = {Figuring {Out} {Gas} \& {Galaxies} in {Enzo} ({FOGGIE}). {II}. {Emission} from the z = 3 {Circumgalactic} {Medium}},
    volume = {896},
    issn = {0004-637X, 1538-4357},
    url = {https://iopscience.iop.org/article/10.3847/1538-4357/ab9310},
    doi = {10.3847/1538-4357/ab9310},
    language = {en},
    number = {2},
    urldate = {2025-05-28},
    journal = {The Astrophysical Journal},
    author = {Corlies, Lauren and Peeples, Molly S. and Tumlinson, Jason and O’Shea, Brian W. and Lehner, Nicolas and Howk, J. Christopher and O’Meara, John M. and Smith, Britton D.},
    month = jun,
    year = {2020},
    pages = {125},
}

@article{simons_figuring_2020,
    title = {Figuring {Out} {Gas} \& {Galaxies} in {Enzo} ({FOGGIE}). {IV}. {The} {Stochasticity} of {Ram} {Pressure} {Stripping} in {Galactic} {Halos}},
    volume = {905},
    issn = {0004-637X, 1538-4357},
    url = {https://iopscience.iop.org/article/10.3847/1538-4357/abc5b8},
    doi = {10.3847/1538-4357/abc5b8},
    language = {en},
    number = {2},
    urldate = {2025-05-28},
    journal = {The Astrophysical Journal},
    author = {Simons, Raymond C. and Peeples, Molly S. and Tumlinson, Jason and O’Shea, Brian W. and Smith, Britton D. and Corlies, Lauren and Lochhaas, Cassandra and Zheng, Yong and Augustin, Ramona and Prasad, Deovrat and Snyder, Gregory F. and Tollerud, Erik},
    month = dec,
    year = {2020},
    pages = {167},
}

@article{wright_figuring_2024,
    title = {Figuring {Out} {Gas} and {Galaxies} in {Enzo} ({FOGGIE}). {VII}. {The} ({Dis})assembly of {Stellar} {Halos}},
    language = {en},
    journal = {The Astrophysical Journal},
    author = {Wright, Anna C and Tumlinson, Jason and Peeples, Molly S and O’Shea, Brian W and Lochhaas, Cassandra and Corlies, Lauren and Smith, Britton D and Binh, Nguyen and Augustin, Ramona and Simons, Raymond C},
    year = {2024},
}

@article{bryan_enzo_2014,
    title = {Enzo: {An} {Adaptive} {Mesh} {Refinement} {Code} for {Astrophysics}},
    volume = {211},
    issn = {0067-0049, 1538-4365},
    shorttitle = {Enzo},
    url = {http://arxiv.org/abs/1307.2265},
    doi = {10.1088/0067-0049/211/2/19},
    number = {2},
    urldate = {2025-07-18},
    journal = {The Astrophysical Journal Supplement Series},
    author = {Bryan, Greg L. and Norman, Michael L. and O'Shea, Brian W. and Abel, Tom and Wise, John H. and Turk, Matthew J. and Reynolds, Daniel R. and Collins, David C. and Wang, Peng and Skillman, Samuel W. and Smith, Britton and Harkness, Robert P. and Bordner, James and Kim, Ji-hoon and Kuhlen, Michael and Xu, Hao and Goldbaum, Nathan and Hummels, Cameron and Kritsuk, Alexei G. and Tasker, Elizabeth and Skory, Stephen and Simpson, Christine M. and Hahn, Oliver and Oishi, Jeffrey S. and So, Geoffrey C. and Zhao, Fen and Cen, Renyue and Li, Yuan},
    month = mar,
    year = {2014},
    note = {arXiv:1307.2265 [astro-ph]},
    pages = {19},
}

@misc{lehner_project_2025,
    title = {Project {AMIGA}: {The} {Inner} {Circumgalactic} {Medium} of {Andromeda} from {Thick} {Disk} to {Halo}},
    shorttitle = {Project {AMIGA}},
    url = {http://arxiv.org/abs/2506.16573},
    doi = {10.48550/arXiv.2506.16573},
    urldate = {2025-07-14},
    publisher = {arXiv},
    author = {Lehner, Nicolas and Howk, J. Christopher and Collins, Lucy and Sameer and Wakker, Bart P. and Augustin, Ramona and Barger, Kathleen A. and Berg, Michelle A. and Bordoloi, Rongmon and Brown, Thomas M. and Cashman, Frances H. and Faucher-Giguère, Claude-André and Fox, Andrew J. and French, David M. and Gilbert, Karoline M. and Guhathakurta, Puragra and O'Meara, John M. and O'Shea, Brian W. and Peeples, Molly S. and Pisano, D. J. and Prochaska, J. Xavier and Stern, Jonathan and Tumlinson, Jason and Werk, Jessica K. and Williams, Benjamin F.},
    month = jun,
    year = {2025},
    note = {arXiv:2506.16573 [astro-ph]},
}

@article{boyd_salsa_2020,
    title = {{SALSA}: {A} {Python} {Package} for {Constructing} {Synthetic} {Quasar} {Absorption} {Line} {Catalogs} from {Astrophysical} {Hydrodynamic} {Simulations}},
    volume = {5},
    issn = {2475-9066},
    shorttitle = {{SALSA}},
    url = {https://joss.theoj.org/papers/10.21105/joss.02581},
    doi = {10.21105/joss.02581},
    language = {en},
    number = {52},
    urldate = {2025-07-18},
    journal = {Journal of Open Source Software},
    author = {Boyd, Brendan I. and Silvia, Devin W. and O'Shea, Brian W. and Tumlinson, Jason and Peeples, Molly S. and Earl, Nicholas},
    month = aug,
    year = {2020},
    pages = {2581},
}

@article{werk_cos-halos_2016,
    title = {The {COS}-{Halos} {Survey}: {Origins} of the {Highly} {Ionized} {Circumgalactic} {Medium} of {Star}-{Forming} {Galaxies}},
    volume = {833},
    issn = {0004-637X},
    shorttitle = {The {COS}-{Halos} {Survey}},
    url = {https://ui.adsabs.harvard.edu/abs/2016ApJ...833...54W},
    doi = {10.3847/1538-4357/833/1/54},
    urldate = {2025-07-23},
    journal = {The Astrophysical Journal},
    author = {Werk, Jessica K. and Prochaska, J. Xavier and Cantalupo, Sebastiano and Fox, Andrew J. and Oppenheimer, Benjamin and Tumlinson, Jason and Tripp, Todd M. and Lehner, Nicolas and McQuinn, Matthew},
    month = dec,
    year = {2016},
    note = {Publisher: IOP
ADS Bibcode: 2016ApJ...833...54W},
    pages = {54},
}

@article{tumlinson_circumgalactic_2017,
    title = {The {Circumgalactic} {Medium}},
    volume = {55},
    issn = {0066-4146},
    url = {https://ui.adsabs.harvard.edu/abs/2017ARA&A..55..389T},
    doi = {10.1146/annurev-astro-091916-055240},
    urldate = {2025-10-01},
    journal = {Annual Review of Astronomy and Astrophysics},
    author = {Tumlinson, Jason and Peeples, Molly S. and Werk, Jessica K.},
    month = aug,
    year = {2017},
    note = {ADS Bibcode: 2017ARA\&A..55..389T},
    pages = {389--432},
}

@misc{ho_kinematics_2025,
    title = {Kinematics of {Circumgalactic} {O} {VI} {Gas} and {Disk} {Rotation} of \$z{\textbackslash}approx0.2\$ {Star}-forming {Galaxies}},
    url = {https://ui.adsabs.harvard.edu/abs/2025arXiv250711664H},
    doi = {10.48550/arXiv.2507.11664},
    urldate = {2025-10-01},
    publisher = {arXiv},
    author = {Ho, Stephanie H. and Martin, Crystal L. and Nateghi, Hasti and Kacprzak, Glenn G. and Stern, Jonathan},
    month = jul,
    year = {2025},
    note = {ADS Bibcode: 2025arXiv250711664H},
}

@article{werk_cos-halos_2014,
    title = {The {COS}-{Halos} {Survey}: {Physical} {Conditions} and {Baryonic} {Mass} in the {Low}-redshift {Circumgalactic} {Medium}},
    volume = {792},
    issn = {0004-637X},
    shorttitle = {The {COS}-{Halos} {Survey}},
    url = {https://ui.adsabs.harvard.edu/abs/2014ApJ...792....8W},
    doi = {10.1088/0004-637X/792/1/8},
    urldate = {2025-10-01},
    journal = {The Astrophysical Journal},
    author = {Werk, Jessica K. and Prochaska, J. Xavier and Tumlinson, Jason and Peeples, Molly S. and Tripp, Todd M. and Fox, Andrew J. and Lehner, Nicolas and Thom, Christopher and O'Meara, John M. and Ford, Amanda Brady and Bordoloi, Rongmon and Katz, Neal and Tejos, Nicolas and Oppenheimer, Benjamin D. and Davé, Romeel and Weinberg, David H.},
    month = sep,
    year = {2014},
    note = {Publisher: IOP
ADS Bibcode: 2014ApJ...792....8W},
    pages = {8},
}

@article{fox_kinematics_2020,
    title = {Kinematics of the {Magellanic} {Stream} and {Implications} for {Its} {Ionization}*},
    volume = {897},
    issn = {0004-637X, 1538-4357},
    url = {https://iopscience.iop.org/article/10.3847/1538-4357/ab92a3},
    doi = {10.3847/1538-4357/ab92a3},
    language = {en},
    number = {1},
    urldate = {2025-05-13},
    journal = {The Astrophysical Journal},
    author = {Fox, Andrew J. and Frazer, Elaine M. and Bland-Hawthorn, Joss and Wakker, Bart P. and Barger, Kathleen A. and Richter, Philipp},
    month = jul,
    year = {2020},
    pages = {23},
}

@article{smith_grackle_2017,
    title = {{GRACKLE}: a chemistry and cooling library for astrophysics},
    volume = {466},
    issn = {0035-8711},
    shorttitle = {{GRACKLE}},
    url = {https://ui.adsabs.harvard.edu/abs/2017MNRAS.466.2217S},
    doi = {10.1093/mnras/stw3291},
    urldate = {2025-10-02},
    journal = {Monthly Notices of the Royal Astronomical Society},
    author = {Smith, Britton D. and Bryan, Greg L. and Glover, Simon C. O. and Goldbaum, Nathan J. and Turk, Matthew J. and Regan, John and Wise, John H. and Schive, Hsi-Yu and Abel, Tom and Emerick, Andrew and O'Shea, Brian W. and Anninos, Peter and Hummels, Cameron B. and Khochfar, Sadegh},
    month = apr,
    year = {2017},
    note = {Publisher: OUP
ADS Bibcode: 2017MNRAS.466.2217S},
    pages = {2217--2234},
}

@article{haardt_radiative_2012,
    title = {Radiative {Transfer} in a {Clumpy} {Universe}. {IV}. {New} {Synthesis} {Models} of the {Cosmic} {UV}/{X}-{Ray} {Background}},
    volume = {746},
    issn = {0004-637X},
    url = {https://ui.adsabs.harvard.edu/abs/2012ApJ...746..125H},
    doi = {10.1088/0004-637X/746/2/125},
    urldate = {2025-10-02},
    journal = {The Astrophysical Journal},
    author = {Haardt, Francesco and Madau, Piero},
    month = feb,
    year = {2012},
    note = {Publisher: IOP
ADS Bibcode: 2012ApJ...746..125H},
    pages = {125},
}

@article{cen_where_2006,
    title = {Where {Are} the {Baryons}? {II}. {Feedback} {Effects}},
    volume = {650},
    issn = {0004-637X},
    shorttitle = {Where {Are} the {Baryons}?},
    url = {https://iopscience.iop.org/article/10.1086/506505/meta},
    doi = {10.1086/506505},
    language = {en},
    number = {2},
    urldate = {2025-10-02},
    journal = {The Astrophysical Journal},
    author = {Cen, Renyue and Ostriker, Jeremiah P.},
    month = oct,
    year = {2006},
    note = {Publisher: IOP Publishing},
    pages = {560},
}

@article{chen_what_2010,
    title = {What {Determines} the {Incidence} and {Extent} of {Mg} {II} {Absorbing} {Gas} {Around} {Galaxies}?},
    volume = {724},
    issn = {0004-637X},
    url = {https://ui.adsabs.harvard.edu/abs/2010ApJ...724L.176C},
    doi = {10.1088/2041-8205/724/2/L176},
    urldate = {2025-10-02},
    journal = {The Astrophysical Journal},
    author = {Chen, Hsiao-Wen and Wild, Vivienne and Tinker, Jeremy L. and Gauthier, Jean-René and Helsby, Jennifer E. and Shectman, Stephen A. and Thompson, Ian B.},
    month = dec,
    year = {2010},
    note = {Publisher: IOP
ADS Bibcode: 2010ApJ...724L.176C},
    pages = {L176--L182},
}

@article{taira_impacts_2025,
    title = {Impacts of the {Metagalactic} {Ultraviolet} {Background} on {Circumgalactic} {Medium} {Absorption} {Systems}},
    volume = {991},
    issn = {0004-637X},
    url = {https://doi.org/10.3847/1538-4357/adfc4e},
    doi = {10.3847/1538-4357/adfc4e},
    language = {en},
    number = {2},
    urldate = {2025-10-03},
    journal = {The Astrophysical Journal},
    author = {Taira, Elias and Kopenhafer, Claire and O’Shea, Brian W. and Manning, Alexis and Fuhrman, Evelyn and Peeples, Molly S. and Tumlinson, Jason and Smith, Britton D.},
    month = sep,
    year = {2025},
    note = {Publisher: The American Astronomical Society},
    pages = {221},
}

@article{strawn_agora_2024,
    title = {The {AGORA} {High}-resolution {Galaxy} {Simulations} {Comparison} {Project}. {VI}. {Similarities} and {Differences} in the {Circumgalactic} {Medium}},
    volume = {962},
    issn = {0004-637X},
    url = {https://ui.adsabs.harvard.edu/abs/2024ApJ...962...29S},
    doi = {10.3847/1538-4357/ad12cb},
    urldate = {2025-07-22},
    journal = {The Astrophysical Journal},
    author = {Strawn, Clayton and Roca-Fàbrega, Santi and Primack, Joel R. and Kim, Ji-Hoon and Genina, Anna and Hausammann, Loic and Kim, Hyeonyong and Lupi, Alessandro and Nagamine, Kentaro and Powell, Johnny W. and Revaz, Yves and Shimizu, Ikkoh and Velázquez, Héctor and Abel, Tom and Ceverino, Daniel and Dong, Bili and Jung, Minyong and Quinn, Thomas R. and Shin, Eun-Jin and Barrow, Kirk S. S. and Dekel, Avishai and Oh, Boon Kiat and Mandelker, Nir and Teyssier, Romain and Hummels, Cameron and Maji, Soumily and Man, Antonio and Mayerhofer, Paul and {The Agora Collaboration}},
    month = feb,
    year = {2024},
    note = {Publisher: IOP
ADS Bibcode: 2024ApJ...962...29S},
    pages = {29},
}

@article{brummel-smith_enzo_2019,
    title = {{ENZO}: {An} {Adaptive} {Mesh} {Refinement} {Code} for {Astrophysics} ({Version} 2.6)},
    volume = {4},
    copyright = {http://creativecommons.org/licenses/by/4.0/},
    issn = {2475-9066},
    shorttitle = {{ENZO}},
    url = {https://joss.theoj.org/papers/10.21105/joss.01636},
    doi = {10.21105/joss.01636},
    language = {en},
    number = {42},
    urldate = {2025-10-11},
    journal = {Journal of Open Source Software},
    author = {Brummel-Smith, Corey and Bryan, Greg and Butsky, Iryna and Corlies, Lauren and Emerick, Andrew and Forbes, John and Fujimoto, Yusuke and Goldbaum, Nathan and Grete, Philipp and Hummels, Cameron and Kim, Ji-hoon and Koh, Daegene and Li, Miao and Li, Yuan and Li, Xinyu and O'Shea, Brian and Peeples, Molly and Regan, John and Salem, Munier and Schmidt, Wolfram and Simpson, Christine and Smith, Britton and Tumlinson, Jason and Turk, Matthew and Wise, John and Abel, Tom and Bordner, James and Cen, Renyue and Collins, David and Crosby, Brian and Edelmann, Philipp and Hahn, Oliver and Harkness, Robert and Harper-Clark, Elizabeth and Kong, Shuo and Kritsuk, Alexei and Kuhlen, Michael and Larrue, James and Lee, Eve and Meece, Greg and Norman, Michael and Oishi, Jeffrey and Paschos, Pascal and Peruta, Carolyn and Razoumov, Alex and Reynolds, Daniel and Silvia, Devin and Skillman, Samuel and Skory, Stephen and So, Geoffrey and Tasker, Elizabeth and Wagner, Rick and Wang, Peng and Xu, Hao and Zhao, Fen},
    month = oct,
    year = {2019},
    note = {TLDR: Corey Brummel-Smith4, Greg Bryan1, 2, Iryna Butsky14, Lauren Corlies5, 6, Andrew Emerick1, 10, John Forbes19, Yusuke Fujimoto34, Nathan J. Goldbaum15, Philipp Grete3, Cameron B. Wise4, Tom Abel24, 25, James Bordner20, Renyue Cen27, David C. Reynolds26, Devin Silvia16, Samuel W. Skillman28},
    pages = {1636},
}

@ARTICLE{Astropy1,
       author = {{Astropy Collaboration} and {Robitaille}, Thomas P. and {Tollerud}, Erik J. and {Greenfield}, Perry and {Droettboom}, Michael and {Bray}, Erik and {Aldcroft}, Tom and {Davis}, Matt and {Ginsburg}, Adam and {Price-Whelan}, Adrian M. and {Kerzendorf}, Wolfgang E. and {Conley}, Alexander and {Crighton}, Neil and {Barbary}, Kyle and {Muna}, Demitri and {Ferguson}, Henry and {Grollier}, Fr{\'e}d{\'e}ric and {Parikh}, Madhura M. and {Nair}, Prasanth H. and {Unther}, Hans M. and {Deil}, Christoph and {Woillez}, Julien and {Conseil}, Simon and {Kramer}, Roban and {Turner}, James E.~H. and {Singer}, Leo and {Fox}, Ryan and {Weaver}, Benjamin A. and {Zabalza}, Victor and {Edwards}, Zachary I. and {Azalee Bostroem}, K. and {Burke}, D.~J. and {Casey}, Andrew R. and {Crawford}, Steven M. and {Dencheva}, Nadia and {Ely}, Justin and {Jenness}, Tim and {Labrie}, Kathleen and {Lim}, Pey Lian and {Pierfederici}, Francesco and {Pontzen}, Andrew and {Ptak}, Andy and {Refsdal}, Brian and {Servillat}, Mathieu and {Streicher}, Ole},
        title = "{Astropy: A community Python package for astronomy}",
      journal = {\aap},
         year = 2013,
        month = oct,
       volume = {558},
          eid = {A33},
        pages = {A33},
          doi = {10.1051/0004-6361/201322068},
archivePrefix = {arXiv},
       eprint = {1307.6212},
 primaryClass = {astro-ph.IM},
       adsurl = {https://ui.adsabs.harvard.edu/abs/2013A&A...558A..33A}
}

@ARTICLE{Astropy2,
       author = {{Astropy Collaboration} and {Price-Whelan}, A.~M. and {Sip{\H{o}}cz}, B.~M. and {G{\"u}nther}, H.~M. and {Lim}, P.~L. and {Crawford}, S.~M. and {Conseil}, S. and {Shupe}, D.~L. and {Craig}, M.~W. and {Dencheva}, N. and {Ginsburg}, A. and {VanderPlas}, J.~T. and {Bradley}, L.~D. and {P{\'e}rez-Su{\'a}rez}, D. and {de Val-Borro}, M. and {Aldcroft}, T.~L. and {Cruz}, K.~L. and {Robitaille}, T.~P. and {Tollerud}, E.~J. and {Ardelean}, C. and {Babej}, T. and {Bach}, Y.~P. and {Bachetti}, M. and {Bakanov}, A.~V. and {Bamford}, S.~P. and {Barentsen}, G. and {Barmby}, P. and {Baumbach}, A. and {Berry}, K.~L. and {Biscani}, F. and {Boquien}, M. and {Bostroem}, K.~A. and {Bouma}, L.~G. and {Brammer}, G.~B. and {Bray}, E.~M. and {Breytenbach}, H. and {Buddelmeijer}, H. and {Burke}, D.~J. and {Calderone}, G. and {Cano Rodr{\'\i}guez}, J.~L. and {Cara}, M. and {Cardoso}, J.~V.~M. and {Cheedella}, S. and {Copin}, Y. and {Corrales}, L. and {Crichton}, D. and {D'Avella}, D. and {Deil}, C. and {Depagne}, {\'E}. and {Dietrich}, J.~P. and {Donath}, A. and {Droettboom}, M. and {Earl}, N. and {Erben}, T. and {Fabbro}, S. and {Ferreira}, L.~A. and {Finethy}, T. and {Fox}, R.~T. and {Garrison}, L.~H. and {Gibbons}, S.~L.~J. and {Goldstein}, D.~A. and {Gommers}, R. and {Greco}, J.~P. and {Greenfield}, P. and {Groener}, A.~M. and {Grollier}, F. and {Hagen}, A. and {Hirst}, P. and {Homeier}, D. and {Horton}, A.~J. and {Hosseinzadeh}, G. and {Hu}, L. and {Hunkeler}, J.~S. and {Ivezi{\'c}}, {\v{Z}}. and {Jain}, A. and {Jenness}, T. and {Kanarek}, G. and {Kendrew}, S. and {Kern}, N.~S. and {Kerzendorf}, W.~E. and {Khvalko}, A. and {King}, J. and {Kirkby}, D. and {Kulkarni}, A.~M. and {Kumar}, A. and {Lee}, A. and {Lenz}, D. and {Littlefair}, S.~P. and {Ma}, Z. and {Macleod}, D.~M. and {Mastropietro}, M. and {McCully}, C. and {Montagnac}, S. and {Morris}, B.~M. and {Mueller}, M. and {Mumford}, S.~J. and {Muna}, D. and {Murphy}, N.~A. and {Nelson}, S. and {Nguyen}, G.~H. and {Ninan}, J.~P. and {N{\"o}the}, M. and {Ogaz}, S. and {Oh}, S. and {Parejko}, J.~K. and {Parley}, N. and {Pascual}, S. and {Patil}, R. and {Patil}, A.~A. and {Plunkett}, A.~L. and {Prochaska}, J.~X. and {Rastogi}, T. and {Reddy Janga}, V. and {Sabater}, J. and {Sakurikar}, P. and {Seifert}, M. and {Sherbert}, L.~E. and {Sherwood-Taylor}, H. and {Shih}, A.~Y. and {Sick}, J. and {Silbiger}, M.~T. and {Singanamalla}, S. and {Singer}, L.~P. and {Sladen}, P.~H. and {Sooley}, K.~A. and {Sornarajah}, S. and {Streicher}, O. and {Teuben}, P. and {Thomas}, S.~W. and {Tremblay}, G.~R. and {Turner}, J.~E.~H. and {Terr{\'o}n}, V. and {van Kerkwijk}, M.~H. and {de la Vega}, A. and {Watkins}, L.~L. and {Weaver}, B.~A. and {Whitmore}, J.~B. and {Woillez}, J. and {Zabalza}, V. and {Astropy Contributors}},
        title = "{The Astropy Project: Building an Open-science Project and Status of the v2.0 Core Package}",
      journal = {\aj},
         year = 2018,
        month = sep,
       volume = {156},
       number = {3},
          eid = {123},
        pages = {123},
          doi = {10.3847/1538-3881/aabc4f},
archivePrefix = {arXiv},
       eprint = {1801.02634},
 primaryClass = {astro-ph.IM},
       adsurl = {https://ui.adsabs.harvard.edu/abs/2018AJ....156..123A}
}

@ARTICLE{Turk2011,
   author = {{Turk}, M.~J. and {Smith}, B.~D. and {Oishi}, J.~S. and {Skory}, S. and
     {Skillman}, S.~W. and {Abel}, T. and {Norman}, M.~L.},
    title = "{yt: A Multi-code Analysis Toolkit for Astrophysical Simulation Data}",
  journal = {The Astrophysical Journal Supplement Series},
archivePrefix = "arXiv",
   eprint = {1011.3514},
 primaryClass = "astro-ph.IM",
     year = 2011,
    month = jan,
   volume = 192,
      eid = {9},
    pages = {9},
      doi = {10.1088/0067-0049/192/1/9},
   adsurl = {https://ui.adsabs.harvard.edu/abs/2011ApJS..192....9T}
}

@ARTICLE{hunter2007,
       author = {{Hunter}, John D.},
        title = "{Matplotlib: A 2D Graphics Environment}",
      journal = {Computing in Science and Engineering},
         year = 2007,
        month = may,
       volume = {9},
       number = {3},
        pages = {90-95},
          doi = {10.1109/MCSE.2007.55},
       adsurl = {https://ui.adsabs.harvard.edu/abs/2007CSE.....9...90H}
}

@article{walt2011numpy,
  title={The NumPy array: a structure for efficient numerical computation},
  author={Walt, St{\'e}fan van der and Colbert, S Chris and Varoquaux, Gael},
  journal={Computing in Science \& Engineering},
  volume={13},
  number={2},
  pages={22--30},
  year={2011},
  publisher={IEEE}
}

@ARTICLE{scipy2020,
       author = {{Virtanen}, Pauli and {Gommers}, Ralf and {Oliphant}, Travis E. and {Haberland}, Matt and {Reddy}, Tyler and {Cournapeau}, David and {Burovski}, Evgeni and {Peterson}, Pearu and {Weckesser}, Warren and {Bright}, Jonathan and {van der Walt}, St{\'e}fan J. and {Brett}, Matthew and {Wilson}, Joshua and {Millman}, K. Jarrod and {Mayorov}, Nikolay and {Nelson}, Andrew R.~J. and {Jones}, Eric and {Kern}, Robert and {Larson}, Eric and {Carey}, C.~J. and {Polat}, {\.I}lhan and {Feng}, Yu and {Moore}, Eric W. and {VanderPlas}, Jake and {Laxalde}, Denis and {Perktold}, Josef and {Cimrman}, Robert and {Henriksen}, Ian and {Quintero}, E.~A. and {Harris}, Charles R. and {Archibald}, Anne M. and {Ribeiro}, Ant{\^o}nio H. and {Pedregosa}, Fabian and {van Mulbregt}, Paul and {SciPy 1. 0 Contributors}},
        title = "{SciPy 1.0: fundamental algorithms for scientific computing in Python}",
      journal = {Nature Methods},
         year = 2020,
        month = feb,
       volume = {17},
        pages = {261-272},
          doi = {10.1038/s41592-019-0686-2},
archivePrefix = {arXiv},
       eprint = {1907.10121},
 primaryClass = {cs.MS},
       adsurl = {https://ui.adsabs.harvard.edu/abs/2020NatMe..17..261V}
}

@article{lehner_evidence_2015,
    title = {{EVIDENCE} {FOR} {A} {MASSIVE}, {EXTENDED} {CIRCUMGALACTIC} {MEDIUM} {AROUND} {THE} {ANDROMEDA} {GALAXY}},
    volume = {804},
    copyright = {http://iopscience.iop.org/info/page/text-and-data-mining},
    issn = {1538-4357},
    url = {https://iopscience.iop.org/article/10.1088/0004-637X/804/2/79},
    doi = {10.1088/0004-637X/804/2/79},
    language = {en},
    number = {2},
    urldate = {2025-11-24},
    journal = {The Astrophysical Journal},
    author = {Lehner, Nicolas and Howk, J. Christopher and Wakker, Bart P.},
    month = may,
    year = {2015},
    pages = {22},
}

@article{lehner_kodiaq-z_2022,
    title = {{KODIAQ}-{Z}: {Metals} and {Baryons} in the {Cool} {Intergalactic} and {Circumgalactic} {Gas} at 2.2 ≲ z ≲ 3.6},
    volume = {936},
    issn = {0004-637X, 1538-4357},
    shorttitle = {{KODIAQ}-{Z}},
    url = {https://iopscience.iop.org/article/10.3847/1538-4357/ac7400},
    doi = {10.3847/1538-4357/ac7400},
    language = {en},
    number = {2},
    urldate = {2025-08-13},
    journal = {The Astrophysical Journal},
    author = {Lehner, Nicolas and Kopenhafer, Claire and O’Meara, John M. and Howk, J. Christopher and Fumagalli, Michele and Prochaska, J. Xavier and Acharyya, Ayan and O’Shea, Brian W. and Peeples, Molly S. and Tumlinson, Jason and Hummels, Cameron B.},
    month = sep,
    year = {2022},
    pages = {156},
}

@article{tripp_highresolution_2008,
    title = {A {High}‐{Resolution} {Survey} of {Low}‐{Redshift} {QSO} {Absorption} {Lines}: {Statistics} and {Physical} {Conditions} of {O} {\textless}span style="font-variant:small-caps;"{\textgreater}vi{\textless}/span{\textgreater} {Absorbers}},
    volume = {177},
    issn = {0067-0049, 1538-4365},
    shorttitle = {A {High}‐{Resolution} {Survey} of {Low}‐{Redshift} {QSO} {Absorption} {Lines}},
    url = {https://iopscience.iop.org/article/10.1086/587486},
    doi = {10.1086/587486},
    language = {en},
    number = {1},
    urldate = {2025-12-02},
    journal = {The Astrophysical Journal Supplement Series},
    author = {Tripp, Todd M. and Sembach, Kenneth R. and Bowen, David V. and Savage, Blair D. and Jenkins, Edward B. and Lehner, Nicolas and Richter, Philipp},
    month = jul,
    year = {2008},
    pages = {39--102},
}

@article{cen_galaxy_1992,
    title = {Galaxy formation and physical bias},
    volume = {399},
    issn = {0004-637X, 1538-4357},
    url = {http://adsabs.harvard.edu/doi/10.1086/186620},
    doi = {10.1086/186620},
    language = {en},
    urldate = {2025-12-02},
    journal = {The Astrophysical Journal},
    author = {Cen, Renyue and Ostriker, Jeremiah P.},
    month = nov,
    year = {1992},
    pages = {L113},
}

@article{smith_nature_2011,
    title = {The {Nature} of the {Warm}/{Hot} {Intergalactic} {Medium}. {I}. {Numerical} {Methods}, {Convergence}, and {O} {VI} {Absorption}},
    volume = {731},
    issn = {0004-637X},
    url = {https://ui.adsabs.harvard.edu/abs/2011ApJ...731....6S},
    doi = {10.1088/0004-637X/731/1/6},
    urldate = {2025-12-02},
    journal = {The Astrophysical Journal},
    author = {Smith, Britton D. and Hallman, Eric J. and Shull, J. Michael and O'Shea, Brian W.},
    month = apr,
    year = {2011},
    note = {Publisher: IOP
ADS Bibcode: 2011ApJ...731....6S},
    pages = {6},
}

@article{helmi_merger_2018,
    title = {The merger that led to the formation of the {Milky} {Way}’s inner stellar halo and thick disk},
    volume = {563},
    issn = {1476-4687},
    url = {https://doi.org/10.1038/s41586-018-0625-x},
    doi = {10.1038/s41586-018-0625-x},
    number = {7729},
    journal = {Nature},
    author = {Helmi, Amina and Babusiaux, Carine and Koppelman, Helmer H. and Massari, Davide and Veljanoski, Jovan and Brown, Anthony G. A.},
    month = nov,
    year = {2018},
    pages = {85--88},
}

@article{lehner_galactic_2014,
    title = {{GALACTIC} {AND} {CIRCUMGALACTIC} {O} {VI} {AND} {ITS} {IMPACT} {ON} {THE} {COSMOLOGICAL} {METAL} {AND} {BARYON} {BUDGETS} {AT} 2 {\textless} \textit{z} ≲ 3.5},
    volume = {788},
    copyright = {http://iopscience.iop.org/info/page/text-and-data-mining},
    issn = {0004-637X, 1538-4357},
    url = {https://iopscience.iop.org/article/10.1088/0004-637X/788/2/119},
    doi = {10.1088/0004-637X/788/2/119},
    language = {en},
    number = {2},
    urldate = {2025-12-02},
    journal = {The Astrophysical Journal},
    author = {Lehner, N. and O'Meara, J. M. and Fox, A. J. and Howk, J. C. and Prochaska, J. X. and Burns, V. and Armstrong, A. A.},
    month = may,
    year = {2014},
    pages = {119},
}

@article{vandevoort_cosmological_2019,
    title = {Cosmological simulations of the circumgalactic medium with 1 kpc resolution: enhanced {H} {\textless}span style="font-variant:small-caps;"{\textgreater}i{\textless}/span{\textgreater} column densities},
    volume = {482},
    copyright = {https://academic.oup.com/journals/pages/open\_access/funder\_policies/chorus/standard\_publication\_model},
    issn = {1745-3925, 1745-3933},
    shorttitle = {Cosmological simulations of the circumgalactic medium with 1 kpc resolution},
    url = {https://academic.oup.com/mnrasl/article/482/1/L85/5126365},
    doi = {10.1093/mnrasl/sly190},
    language = {en},
    number = {1},
    urldate = {2025-12-16},
    journal = {Monthly Notices of the Royal Astronomical Society: Letters},
    author = {van de Voort, Freeke and Springel, Volker and Mandelker, Nir and van den Bosch, Frank C and Pakmor, Rüdiger},
    month = jan,
    year = {2019},
    pages = {L85--L89},
}

@article{hummels_impact_2019,
    title = {The {Impact} of {Enhanced} {Halo} {Resolution} on the {Simulated} {Circumgalactic} {Medium}},
    volume = {882},
    issn = {1538-4357},
    url = {https://iopscience.iop.org/article/10.3847/1538-4357/ab378f},
    doi = {10.3847/1538-4357/ab378f},
    language = {en},
    number = {2},
    urldate = {2025-12-16},
    journal = {The Astrophysical Journal},
    author = {Hummels, Cameron B. and Smith, Britton D. and Hopkins, Philip F. and O’Shea, Brian W. and Silvia, Devin W. and Werk, Jessica K. and Lehner, Nicolas and Wise, John H. and Collins, David C. and Butsky, Iryna S.},
    month = sep,
    year = {2019},
    pages = {156},
}

@article{suresh_zooming_2019,
    title = {Zooming in on accretion – {II}. {Cold} circumgalactic gas simulated with a super-{Lagrangian} refinement scheme},
    volume = {483},
    copyright = {https://academic.oup.com/journals/pages/open\_access/funder\_policies/chorus/standard\_publication\_model},
    issn = {0035-8711, 1365-2966},
    url = {https://academic.oup.com/mnras/article/483/3/4040/5251841},
    doi = {10.1093/mnras/sty3402},
    language = {en},
    number = {3},
    urldate = {2025-12-16},
    journal = {Monthly Notices of the Royal Astronomical Society},
    author = {Suresh, Joshua and Nelson, Dylan and Genel, Shy and Rubin, Kate H R and Hernquist, Lars},
    month = mar,
    year = {2019},
    pages = {4040--4059},
}

@article{stocke_characterizing_2013,
    title = {{CHARACTERIZING} {THE} {CIRCUMGALACTIC} {MEDIUM} {OF} {NEARBY} {GALAXIES} {WITH} \textit{{HST}} /{COS} {AND} \textit{{HST}} /{STIS} {ABSORPTION}-{LINE} {SPECTROSCOPY}},
    volume = {763},
    copyright = {http://iopscience.iop.org/info/page/text-and-data-mining},
    issn = {0004-637X, 1538-4357},
    url = {https://iopscience.iop.org/article/10.1088/0004-637X/763/2/148},
    doi = {10.1088/0004-637X/763/2/148},
    language = {en},
    number = {2},
    urldate = {2025-12-16},
    journal = {The Astrophysical Journal},
    author = {Stocke, John T. and Keeney, Brian A. and Danforth, Charles W. and Shull, J. Michael and Froning, Cynthia S. and Green, James C. and Penton, Steven V. and Savage, Blair D.},
    month = jan,
    year = {2013},
    pages = {148},
}

@article{emerick_simulating_2019,
    title = {Simulating an isolated dwarf galaxy with multichannel feedback and chemical yields from individual stars},
    volume = {482},
    copyright = {https://academic.oup.com/journals/pages/open\_access/funder\_policies/chorus/standard\_publication\_model},
    issn = {0035-8711, 1365-2966},
    url = {https://academic.oup.com/mnras/article/482/1/1304/5115580},
    doi = {10.1093/mnras/sty2689},
    language = {en},
    number = {1},
    urldate = {2025-12-17},
    journal = {Monthly Notices of the Royal Astronomical Society},
    author = {Emerick, Andrew and Bryan, Greg L and Mac Low, Mordecai-Mark},
    month = jan,
    year = {2019},
    pages = {1304--1329},
}

@misc{sameer_cos_2024,
    title = {The {COS} {CGM} {Compendium} {V}: {The} {Dichotomy} of {OVI} {Associated} with {Low}- and {High}-{Metallicity} {Cool} {Gas} at z {\textless} 1},
    shorttitle = {The {COS} {CGM} {Compendium} {V}},
    url = {http://arxiv.org/abs/2403.02374},
    doi = {10.48550/arXiv.2403.02374},
    language = {en},
    urldate = {2025-08-19},
    publisher = {arXiv},
    author = {Sameer and Lehner, Nicolas and Howk, J. Christopher and Fox, Andrew J. and O'Meara, John M. and Oppenheimer, Benjamin D.},
    month = sep,
    year = {2024},
    note = {arXiv:2403.02374 [astro-ph]},
}

@article{chen_circumgalactic_2025,
    title = {The {Circumgalactic} {Medium} {Traced} by {Mg} {\textless}span style="font-variant:small-caps;"{\textgreater}ii{\textless}/span{\textgreater} {Absorption} with {DESI}: {Dependence} on {Galaxy} {Stellar} {Mass}, {Star} {Formation} {Rate}, and {Azimuthal} {Angle}},
    volume = {981},
    issn = {0004-637X, 1538-4357},
    shorttitle = {The {Circumgalactic} {Medium} {Traced} by {Mg} {\textless}span style="font-variant},
    url = {https://iopscience.iop.org/article/10.3847/1538-4357/ada942},
    doi = {10.3847/1538-4357/ada942},
    language = {en},
    number = {1},
    urldate = {2025-12-17},
    journal = {The Astrophysical Journal},
    author = {Chen, Zeyu and Wang, Enci and Zou, Hu and Zou, Siwei and Gao, Yang and Wang, Huiyuan and Yu, Haoran and Jia, Cheng and Li, Haixin and Ma, Chengyu and Yao, Yao and Ding, Weiyu and Zhu, Runyu},
    month = mar,
    year = {2025},
    pages = {81},
}

@article{cherrey_muse_2025,
    title = {{MusE} {GAs} {FLOw} and {Wind} ({MEGAFLOW}): {XIII}. {Cool} gas traced by {Mg} {II} around isolated galaxies},
    volume = {694},
    copyright = {https://creativecommons.org/licenses/by/4.0},
    issn = {0004-6361, 1432-0746},
    shorttitle = {{MusE} {GAs} {FLOw} and {Wind} ({MEGAFLOW})},
    url = {https://www.aanda.org/10.1051/0004-6361/202451165},
    doi = {10.1051/0004-6361/202451165},
    language = {en},
    urldate = {2025-12-17},
    journal = {Astronomy \& Astrophysics},
    author = {Cherrey, Maxime and Bouché, Nicolas F. and Zabl, Johannes and Schroetter, Ilane and Wendt, Martin and Langan, Ivanna and Schaye, Joop and Wisotzki, Lutz and Guo, Yucheng and Pessa, Ismael},
    month = feb,
    year = {2025},
    pages = {A117},
}

@article{dutta_musequbes_2025,
    title = {{MUSEQuBES}: {The} {Column} {Density}, {Covering} {Fraction}, and {Mass} of {O} {VI}-bearing {Gas} in and {Around} {Low}-redshift {Galaxies}},
    volume = {985},
    issn = {0004-637X},
    shorttitle = {{MUSEQuBES}},
    url = {https://ui.adsabs.harvard.edu/abs/2025ApJ...985...44D},
    doi = {10.3847/1538-4357/adc922},
    urldate = {2025-07-24},
    journal = {The Astrophysical Journal},
    author = {Dutta, Sayak and Muzahid, Sowgat and Schaye, Joop and Bouché, Nicolas F. and Cantalupo, Sebastiano and Chen, Hsiao-Wen and Johnson, Sean},
    month = may,
    year = {2025},
    note = {Publisher: IOP
ADS Bibcode: 2025ApJ...985...44D},
    pages = {44},
}

@article{peroux_cosmic_2020,
    title = {The {Cosmic} {Baryon} and {Metal} {Cycles}},
    volume = {58},
    issn = {0066-4146, 1545-4282},
    url = {https://www.annualreviews.org/doi/10.1146/annurev-astro-021820-120014},
    doi = {10.1146/annurev-astro-021820-120014},
    language = {en},
    number = {1},
    urldate = {2025-12-17},
    journal = {Annual Review of Astronomy and Astrophysics},
    author = {Péroux, Céline and Howk, J. Christopher},
    month = aug,
    year = {2020},
    pages = {363--406},
}

@article{vandevoort_effect_2021,
    title = {The effect of magnetic fields on properties of the circumgalactic medium},
    volume = {501},
    copyright = {https://academic.oup.com/journals/pages/open\_access/funder\_policies/chorus/standard\_publication\_model},
    issn = {0035-8711, 1365-2966},
    url = {https://academic.oup.com/mnras/article/501/4/4888/6066534},
    doi = {10.1093/mnras/staa3938},
    language = {en},
    number = {4},
    urldate = {2025-12-17},
    journal = {Monthly Notices of the Royal Astronomical Society},
    author = {van de Voort, Freeke and Bieri, Rebekka and Pakmor, Rüdiger and Gómez, Facundo A and Grand, Robert J J and Marinacci, Federico},
    month = jan,
    year = {2021},
    pages = {4888--4902},
}

@article{weber_crexit_2025,
    title = {{CRexit}: {How} different cosmic ray transport modes affect thermal instability in the circumgalactic medium},
    volume = {698},
    copyright = {https://creativecommons.org/licenses/by/4.0},
    issn = {0004-6361, 1432-0746},
    shorttitle = {{CRexit}},
    url = {https://www.aanda.org/10.1051/0004-6361/202553954},
    doi = {10.1051/0004-6361/202553954},
    language = {en},
    urldate = {2025-12-17},
    journal = {Astronomy \& Astrophysics},
    author = {Weber, M. and Thomas, T. and Pfrommer, C. and Pakmor, R.},
    month = jun,
    year = {2025},
    pages = {A125},
}

@article{butsky_impact_2022,
    title = {The {Impact} of {Cosmic} {Rays} on the {Kinematics} of the {Circumgalactic} {Medium}},
    volume = {935},
    issn = {0004-637X, 1538-4357},
    url = {https://iopscience.iop.org/article/10.3847/1538-4357/ac7ebd},
    doi = {10.3847/1538-4357/ac7ebd},
    language = {en},
    number = {2},
    urldate = {2025-12-17},
    journal = {The Astrophysical Journal},
    author = {Butsky, Iryna S. and Werk, Jessica K. and Tchernyshyov, Kirill and Fielding, Drummond B. and Breneman, Joseph and Piacitelli, Daniel R. and Quinn, Thomas R. and Sanchez, N. Nicole and Cruz, Akaxia and Hummels, Cameron B. and Burchett, Joseph N. and Tremmel, Michael},
    month = aug,
    year = {2022},
    pages = {69},
}

@ARTICLE{2024ApJ...965..100Q,
       author = {{Qu}, Zhijie and {Pan}, Zeyang and {Bregman}, Joel N. and {Liu}, Jifeng},
        title = "{The XMM-Newton Line Emission Analysis Program (X-LEAP). II. The Multiscale Temperature Structures in the Milky Way Hot Gas}",
      journal = {\apj},
         year = 2024,
        month = apr,
       volume = {965},
       number = {2},
          eid = {100},
        pages = {100},
          doi = {10.3847/1538-4357/ad31a0},
archivePrefix = {arXiv},
       eprint = {2403.05664},
 primaryClass = {astro-ph.GA},
       adsurl = {https://ui.adsabs.harvard.edu/abs/2024ApJ...965..100Q}
}

@article{lehner_project_2020,
    title = {Project {AMIGA}: {The} {Circumgalactic} {Medium} of {Andromeda}*},
    volume = {900},
    issn = {0004-637X, 1538-4357},
    shorttitle = {Project {AMIGA}},
    url = {https://iopscience.iop.org/article/10.3847/1538-4357/aba49c},
    doi = {10.3847/1538-4357/aba49c},
    language = {en},
    number = {1},
    urldate = {2025-12-22},
    journal = {The Astrophysical Journal},
    author = {Lehner, Nicolas and Berek, Samantha C. and Howk, J. Christopher and Wakker, Bart P. and Tumlinson, Jason and Jenkins, Edward B. and Prochaska, J. Xavier and Augustin, Ramona and Ji, Suoqing and Faucher-Giguère, Claude-André and Hafen, Zachary and Peeples, Molly S. and Barger, Kat A. and Berg, Michelle A. and Bordoloi, Rongmon and Brown, Thomas M. and Fox, Andrew J. and Gilbert, Karoline M. and Guhathakurta, Puragra and Kalirai, Jason S. and Lockman, Felix J. and O’Meara, John M. and Pisano, D. J. and Ribaudo, Joseph and Werk, Jessica K.},
    month = sep,
    year = {2020},
    pages = {9},
}

@article{qu_cosmic_2024,
    title = {The {Cosmic} {Ultraviolet} {Baryon} {Survey} ({CUBS}). {VII}. {On} the {Warm}-hot {Circumgalactic} {Medium} {Probed} by {O} vi and {Ne} viii at 0.4 ≲ z ≲ 0.7},
    volume = {968},
    issn = {0004-637X, 1538-4357},
    url = {https://iopscience.iop.org/article/10.3847/1538-4357/ad410b},
    doi = {10.3847/1538-4357/ad410b},
    language = {en},
    number = {1},
    urldate = {2025-12-22},
    journal = {The Astrophysical Journal},
    author = {Qu, Zhijie and Chen, Hsiao-Wen and Johnson, Sean D. and Rudie, Gwen C. and Zahedy, Fakhri S. and DePalma, David and Schaye, Joop and Boettcher, Erin T. and Cantalupo, Sebastiano and Chen, Mandy C. and Faucher-Giguère, Claude-André and Li, Jennifer I-Hsiu and Mulchaey, John S. and Petitjean, Patrick and Rafelski, Marc},
    month = jun,
    year = {2024},
    pages = {8},
}

@article{marra_using_2021,
    title = {Using cosmological simulations and synthetic absorption spectra to assess the accuracy of observationally derived {CGM} metallicities},
    volume = {508},
    issn = {0035-8711},
    url = {https://doi.org/10.1093/mnras/stab2896},
    doi = {10.1093/mnras/stab2896},
    number = {4},
    journal = {Monthly Notices of the Royal Astronomical Society},
    author = {Marra, Rachel and Churchill, Christopher W and Doughty, Caitlin and Kacprzak, Glenn G and Charlton, Jane and , Sameer and Nielsen, Nikole M and Ceverino, Daniel and Trujillo-Gomez, Sebastian},
    month = oct,
    year = {2021},
    note = {\_eprint: https://academic.oup.com/mnras/article-pdf/508/4/4938/40873520/stab2896.pdf},
    pages = {4938--4951},
}

@article{marra_examining_2023,
    title = {Examining quasar absorption-line analysis methods: the tension between simulations and observational assumptions key to modelling clouds},
    volume = {527},
    issn = {0035-8711},
    url = {https://doi.org/10.1093/mnras/stad3735},
    doi = {10.1093/mnras/stad3735},
    number = {4},
    journal = {Monthly Notices of the Royal Astronomical Society},
    author = {Marra, Rachel and Churchill, Christopher W and Kacprzak, Glenn G and Nielsen, Nikole M and Trujillo-Gomez, Sebastian and Lewis, Emmy A},
    month = dec,
    year = {2023},
    note = {\_eprint: https://academic.oup.com/mnras/article-pdf/527/4/10522/54943304/stad3735.pdf},
    pages = {10522--10537},
}

@misc{chatgpt,
  author       = {{OpenAI}},
  title        = {ChatGPT},
  year         = {2025},
  howpublished = {\url{https://chat.openai.com/}},
  note         = {Large language model},
  urldate      = {2025-01-01}
}

@misc{claude_opus,
  author       = {{Anthropic}},
  title        = {Claude (Opus family)},
  year         = {2025},
  howpublished = {\url{https://www.anthropic.com/}},
  note         = {Accessed via Claude interface},
  urldate      = {2025-01-01}
}

@ARTICLE{2025ApJ...993...52A,
       author = {{Augustin}, Ramona and {Tumlinson}, Jason and {Peeples}, Molly S. and {O'Shea}, Brian W. and {Smith}, Britton D. and {Lochhaas}, Cassandra and {Wright}, Anna C. and {Acharyya}, Ayan and {Werk}, Jessica K. and {Lehner}, Nicolas and {Corlies}, Lauren and {Simons}, Raymond C. and {Howk}, J. Christopher and {O'Meara}, John M.},
        title = "{FOGGIE. X. Characterizing the Small-scale Structure of the Circumgalactic Medium and Its Imprint on Observables}",
      journal = {\apj},
         year = 2025,
        month = nov,
       volume = {993},
       number = {1},
          eid = {52},
        pages = {52},
          doi = {10.3847/1538-4357/ae0462},
archivePrefix = {arXiv},
       eprint = {2501.06551},
 primaryClass = {astro-ph.GA},
       adsurl = {https://ui.adsabs.harvard.edu/abs/2025ApJ...993...52A}
}
\bibliographystyle{aasjournal}

\end{document}